\documentclass[journal,twocolumn,web]{ieeecolor}
\let\theoremstyle\relax

\usepackage{amsthm}
\theoremstyle{plain}
\newtheorem{theorem}{Theorem}[section]
\newtheorem{lemma}[theorem]{Lemma}

\theoremstyle{definition}
\newtheorem{definition}[theorem]{Definition}
\theoremstyle{remark}
\newtheorem*{remark}{Remark}
\usepackage{generic}
\usepackage{cite}
\usepackage{amsmath,amssymb,amsfonts}
\usepackage{graphicx}
\usepackage{subfig}
\usepackage{color}

\usepackage{multirow}
\usepackage[ruled,vlined]{algorithm2e}
\usepackage{cuted}

\def\BibTeX{{\rm B\kern-.05em{\sc i\kern-.025em b}\kern-.08em
    T\kern-.1667em\lower.7ex\hbox{E}\kern-.125emX}}
\begin{document}
\title{A Multivariate Non-Gaussian Bayesian Filter Using Power Moments}
\author{Guangyu Wu, \IEEEmembership{Student Member, IEEE},
and Anders Lindquist, \IEEEmembership{Life Fellow, IEEE}
\thanks{
}
\thanks{Guangyu Wu is with Department of Automation, Shanghai Jiao Tong University, Shanghai, China. (e-mail: chinarustin@sjtu.edu.cn).}
\thanks{Anders Lindquist is with Department of Automation and School of Mathematical Sciences, Shanghai Jiao Tong University, Shanghai, China. (e-mail: alq@kth.se).}}

\maketitle

\begin{abstract}
In this paper, we extend our results on the univariate non-Gaussian Bayesian filter using power moments \cite{wu2023non} to the multivariate systems, which can be either linear or nonlinear. Doing this introduces several challenging problems, for example a positive parametrization of the density surrogate, which is not only a problem of filter design, but also one of the multiple dimensional Hamburger moment problem. We propose a parametrization of the density surrogate with the proofs to its existence, Positivstellensatz and uniqueness. Based on it, we analyze the errors of moments of the density estimates by the proposed density surrogate. A discussion on continuous and discrete treatments to the non-Gaussian Bayesian filtering problem is proposed to motivate the research on continuous parametrization of the system state. Simulation results are given to validate our proposed filter. To the best of our knowledge, the proposed filter is the first one implementing the multivariate Bayesian filter with the system state parameterized as a continuous function, which only requires the true states being Lebesgue integrable with first several orders of power moments being finite.
\end{abstract}


\section{Introduction}
\label{Introduction}
\IEEEPARstart{T}he Bayesian filter provides a unified recursive approach for nonlinear filtering problems. One of the first exploration of iterative Bayesian estimation is found in Ho and Lee’s paper \cite{ho1964bayesian}, where principle and procedure of Bayesian filtering are specified. Sprangins \cite{spragins1965note} discussed the iterative application of Bayes rule to sequential parameter estimation. Lin and Yau \cite{lin1967bayesian} and Chien and Fu \cite{chien1967bayesian} discussed Bayesian approach to optimization of adaptive systems. The Bayesian filter consists of an iterative measurement-time update process, although sometimes interpreted by different names. In the time-update step, the one-step ahead prediction of state is calculated by the system equation; in the measurement-update step, the correction to the estimate of state according to the current observation is calculated by the observation equation.

However, the Bayesian filter is more a framework for nonlinear non-Gaussian filtering. The specific filters are indeed implementations of the Bayesian filtering framework. When we are confronted with a nonlinear filtering problem, it is always infeasible for us to obtain an optimal or analytic solution and be content with a suboptimal solution to the Bayesian filter \cite{kushner1967approximations, arasaratnam2009cubature}. Mathematically speaking, the difficulty lies in the intractability of the convolution operation in the time update step. There are three main approaches to treat this problem.

We note that when the densities are assumed to be Gaussians or mixtures of Gaussians, the convolution is tractable. When all the densities involved in the filtering process are constrained to be Gaussian, the filtering process is essentially the Kalman filter \cite{kalman1960new, kalman1961new} with the system and observation equations being linear. Furthermore, its numerous variants were proposed as implementations of this approach, including the extended Kalman filter (EKF) \cite{anderson2012optimal}, the central-difference Kalman filter (CDKF) \cite{schei1997finite, norgaard2000new}, the unscented Kalman filter (UKF) \cite{julier2000new}, and the quadrature Kalman filter (QKF) \cite{ito2000gaussian}. From the perspective of Gaussian mixtures, a smart-sensor-based method was developed to deal with Gaussian mixtures with an exponentially increasing number of Gaussians and design optimal estimation in the pioneering work \cite{lin2017optimal} on UDP-like systems. With the strict constraint on the densities all being Gaussian or a mixture of them, it is feasible for the estimators above to obtain a closed form of solution to the convolution operation. 

However in modern applications, the state of the system and the noises don't always follow the Gaussian distribution. In the filtering problem in econometrics, for example in the analysis of financial time series, the distributions of the noises have “heavy tails", where the Gaussian distribution doesn't apply. Modes are viewed as the central tendencies of a distribution in statistics. When the probability density of the state is multi-modal, it is obviously infeasible to estimate it with a Gaussian distribution. Confronted with the numerous cases where the first approach doesn't apply, people have been seeking for other methods to treat the intractable convolution.

In the second approach, we give up seeking an analytic solution to the convolution. Instead, we attempt to approximate the intractable convolution, where no explicit assumption on the density form is needed. We note that by assuming the densities to have a discrete form of function, we are able to write the intractable prior density as a probability mass function, which is supported on finitely many discrete points we choose on the domain $\mathbb{R}^{d}$. There are numerous such methods, including mulitgrid method and point-mass approximation \cite{cai1995adaptive, bucy1971digital} and Monte Carlo sampling approximation \cite{handschin1969monte, christian1999monte}. These nonparametric methods impose no prior constraints on the density functions, which seem to enjoy the maximum flexibility. However the tradeoff is also very severe: quite a bunch of probability values at discrete states need to be stored and the continuity of the density is sacrificed. It means that when given an arbitrary state, we are not always able to obtain its value of probability. Meanwhile, optional resampling is widely used in the filters to avoid depletion of particles with small probability values. To design a strategy of when and how to perform the resampling operation isn't a trivial task. Moreover, the particle filter suffers from the curse of dimensionality, which requires the filter to store massive particles with the increase of dimension. In some applications, we only consider the states with significant values of probability; however the states of small values of probability are extremely important, e.g. in financial applications. In conclusion, the discrete methods for density characterization are intrinsically infeasible in tackling the problem where the states with less significant values of probability still have dominant impact on the filtering problem, such as heavy-tailed filtering \cite{roth2013student, zhu2021novel}.

Meanwhile, several numerical methods have been proposed to obtain an analytic solution to the convolution in the time-update step, including Gaussian/Laplace approximation \cite{mackay1998choice}, iterative quadrature \cite{freitas1999bayesian, kushner2000nonlinear}, Gaussian sum approximation \cite{sorenson1971recursive, alspach1972nonlinear} and state-space calculus \cite{hanzon2001state}. The variational Bayesian filter treats the filtering problem with time-varying measurement noise parameters \cite{vsmidl2006variational, sarkka2013non, csenoz2021variational}. It parameterizes the posterior density as a product of a Gaussian and an Inverse-Gamma distribution with the latter one being the conjugate prior distribution for the variance of a Gaussian distribution. These are the parametric methods for parameterizing the probability density function in a continuous form. However, the flexibility of these methods is limited. Some of the methods listed above are either not able to be extended to the multivariate case, or are not able to treat the multi-modal prior density. Therefore it is infeasible to apply these methods to a wide range of real applications, where there are numerous density functions which don't fall within the prescribed function classes. Moreover, quantitative approximation performance analyses, e.g. an error upper bound of estimation have not been derived for the existing methods yet. At the same time, the Bayesian filter is a recursive algorithm, which means that the estimation errors in the previous steps will have cumulative effects on the estimates of later steps. However with the estimation algorithms listed above, the cumulative errors are problematic to analyze, which severely decreases the value of these methods in practical use. A multivariate non-Gaussian Bayesian filter with the state estimation parameterized as an analytic function, where there is no constraint on the feasible classes of the prior density function, is extraordinarily desired by the researchers on stochastic filtering. 

In this paper, we attempt to treat the multivariate non-Gaussian Bayesian filtering problem. We first formulate the multivariate non-Gaussian Bayesian filtering problem in Section 2. Then we propose to use the higher order moments to characterize the density function. The multivariate density surrogate is also defined. A formal definition of the truncated Hamburger moment problem is given, and the solution to it is proved to exist in Section 3. A construction of the density surrogate, i.e., parametrization of the density function is proposed in Section 4. In doing this, we follow the Kullback-Leibler optimization scheme of  \cite{georgiou2003kullback}. Then a novel Positivstellensatz is proposed to ensure the positiveness of the multivariate density surrogate. There follows the proof of the uniqueness of the solution to the optimization problem. The map from the parameters of the density surrogate to the power moments is proved to be homeomorphic, which ensures that the gradient-based optimization algorithms can be applied to determining the parameters of the density surrogate. In Section 5, given that the prior is a sub-Gaussian distribution, the estimated moments are proved to be asymptotically unbiased from the true ones using the density surrogate. And by selecting a sufficiently large order, using the density surrogate will not bring significant cumulative errors to the moment estimation of the subsequent filtering steps. Furthermore, an analytic error analysis of the multivariate Bayesian filter, i.e., an upper bound of approximation error of the multivariate prior density is proposed in Section 6, which has not yet been done for the multivariate Bayesian filters with no prior constraints on the classes of the densities. Simulation results on estimating different classes of multivariate density functions, including heavy-tailed ones, are given in Section 7. A simulation of the proposed filter on a robot localization problem is also carried out to validate the proposed filter, with a comparison to the particle filter and the unscented Kalman filter.

\section{Problem formulation}
\label{ProblemFormulation}
In this paper, we consider the non-Gaussian multivariate filtering problem for the following system.
\begin{equation}
\label{System}
    \begin{aligned}
    x_{t+1} &=f_{t}\left( x_{t}\right)+\eta_{t} \\
    y_{t} &=h_{t}\left( x_{t}\right)+\epsilon_{t}
\end{aligned}    
\end{equation}
$t=0,1,2, \cdots$. The state $x_{t}$ is a random vector defined on $\mathbb{R}^{d}$ endowed with its Borel sigma algebra of events. Let the system function be $f_{t}: \mathbb{R}^{d} \mapsto \mathbb{R}^{d}$ and the observation function be $h_{t}: \mathbb{R}^{d} \mapsto \mathbb{R}^{d}$.

The system noise $\eta_{t}$ and observation noise $\epsilon_{t}$ are random vectors defined on $\mathbb{R}^{d}$, of which the distributions are denoted as $\rho_{\eta_{t}}(x_{t})$ and $\rho_{\epsilon_{t}}(x_{t})$ respectively. The distribution of the system noise $\eta_{t}$ can be either a Lebesgue integrable function or a probability mass function. The distribution of the observation noise $\epsilon_{t}$ is assumed to be a Lebesgue integrable function. Due to the very loose constraints on the noises, we are able to treat noises of which the density functions are not smooth, such as Laplacian distributions; or even those are not continuous. The noises are assumed to be independent from each other. The probability density function of the noise random vectors is non-Gaussian if assumed continuous. We note that all the densities are multivariate.

We adopt the Bayesian filter as used in \cite{hanzon2001state} and extend it to the multivariate case. Denoting the  collection of observations $y_{t}, y_{t-1}, \cdots, y_{0}$ as $\mathcal{Y}_{t}$, the conditional densities of the measurement and time updates are given by the following

\noindent \textbf{Measurement update}: For $t=0$,
\begin{equation}
\begin{aligned}
    \rho_{x_{0} \mid \mathcal{Y}_{0}}(x)& =\frac{\rho_{y_{0} \mid x_{0}}\left(y_{0}\right) \rho_{x_{0}}(x)}{\int_{\mathbb{R}^{d}}\rho_{y_{0} \mid x_{0}}\left(y_{0}\right) \rho_{x_{0}}(x)dx} \\
    & =\frac{\rho_{\epsilon_{0}}\left(y_{0}-h_{0} (x)\right) \rho_{x_{0}}(x)}{\int_{\mathbb{R}^{d}}\rho_{\epsilon_{0}}\left(y_{0}-h_{0}(x)\right) \rho_{x_{0}}(x)dx};
\label{Update1}
\end{aligned}
\end{equation}
for $t \geq 1$,
\begin{equation}
\begin{aligned}
\rho_{x_{t} \mid \mathcal{Y}_{t}}(x) & =\frac{\rho_{y_{t} \mid x_{t}}\left(y_{t}\right) \rho_{x_{t} \mid \mathcal{Y}_{t-1}}(x)}{\int_{\mathbb{R}^{d}}\rho_{y_{t} \mid x_{t}}\left(y_{t}\right) \rho_{x_{t} \mid \mathcal{Y}_{t-1}}(x)dx}\\
& =\frac{\rho_{\epsilon_{t}}\left(y_{t}-h_{t}(x)\right) \rho_{x_{t} \mid \mathcal{Y}_{t-1}}(x)}{\int_{\mathbb{R}^{d}}\rho_{\epsilon_{t}}\left(y_{t}-h_{t}(x)\right) \rho_{x_{t} \mid \mathcal{Y}_{t-1}}(x)dx}, x \in \mathbb{R}^{d}.
\label{Update2}
\end{aligned}
\end{equation}

\noindent \textbf{Time update}: For $t \geq 0$,
\begin{equation}
\begin{aligned}
    \rho_{x_{t+1} \mid \mathcal{Y}_{t}}(x) & =\left(\rho_{f_{t} (x_{t}) \mid \mathcal{Y}_{t}} * \rho_{\eta_{t}}\right)(x)\\
    & =\int_{\mathbb{R}^{d}} \rho_{x_{t} \mid \mathcal{Y}_{t}}\left( f^{-1}_{t}(\varepsilon) \right) \rho_{\eta_{t}}(x-\varepsilon) d\varepsilon.
\label{Prediction}
\end{aligned}
\end{equation}

$\rho_{x_{t+1} \mid \mathcal{Y}_{t}}$ is the prior of each filtering step $t+1$. In the following sections of this paper, "prior" refers to $\rho_{x_{t+1} \mid \mathcal{Y}_{t}}$, if not specified otherwise. The measurement update (\ref{Update2}) is a multiplication of the densities, which is a straightforward calculation. However it is always infeasible to obtain an analytic form of the prior in (\ref{Prediction}) when the densities are not Gaussian. 

Due to the intractability of the convolution, we naturally consider approximating $\rho_{x_{t+1} \mid \mathcal{Y}_{t}}$. Inspired by the method of moments \cite{hall2004generalized}, we propose to use the truncated power moments to estimate $\rho_{x_{t+1} \mid \mathcal{Y}_{t}}$. Denote the nonnegative integers as $\mathbb{N}_{0}$. Let $\kappa=\left(k_{1}, \cdots, k_{d}\right) \text { with } k_{j} \in \mathbb{N}_{0}$. Then we are able to write the power moments in the form of linear functional
\begin{equation}
    L^{\rho}\left( \kappa \right) = \int_{\mathbb{R}^{d}}x_{1}^{k_{1}} x_{2}^{k_{2}} \cdots x_{d}^{k_{d}}\rho(x)dx, \quad \kappa \in \mathcal{J}_{2n}
\end{equation}
where
$$
    \mathcal{J}_{2n} := \{ \kappa=\left(k_{1}, \cdots, k_{d}\right) \mid k_{i} \in \mathbb{N}_{0}, k_{i} \leq 2n, i = 1, \cdots, d \}.
$$

Then we can calculate the power moments of $\rho_{x_{t+1} \mid \mathcal{Y}_{t}}$ as
\begin{equation}
\begin{aligned}
    & L^{\rho_{t+1|t}}\left( \kappa \right)\\
    = & \int_{\mathbb{R}^{d}} x_{ 1}^{k_{1}} x_{2}^{k_{2}} \cdots x_{d}^{k_{d}} \cdot \rho_{x_{t+1} \mid \mathcal{Y}_{t}}(x) dx\\
    = & \int_{\mathbb{R}^{d}} x_{ 1}^{k_{1}} x_{2}^{k_{2}} \cdots x_{d}^{k_{d}} \int_{\mathbb{R}^{d}} \rho_{x_{t} \mid \mathcal{Y}_{t}}\left(f^{-1}_{t}(\varepsilon)\right) \rho_{\eta_{t}}(x-\varepsilon) d\varepsilon dx\\
    = & \int_{\mathbb{R}^{d}} \int_{\mathbb{R}^{d}} x_{ 1}^{k_{1}} x_{2}^{k_{2}} \cdots x_{d}^{k_{d}}  \rho_{x_{t} \mid \mathcal{Y}_{t}}\left(f^{-1}_{t}(\varepsilon)\right) \rho_{\eta_{t}}(x-\varepsilon) d\varepsilon dx\\
    = & \int_{\mathbb{R}^{d}} \rho_{x_{t} \mid \mathcal{Y}_{t}}\left(f^{-1}_{t}(\varepsilon)\right)\int_{\mathbb{R}^{d}} x_{ 1}^{k_{1}} x_{2}^{k_{2}} \cdots x_{d}^{k_{d}} \rho_{\eta_{t}}(x-\varepsilon) dx d\varepsilon.
\end{aligned}
\label{MomentUpdate}
\end{equation}

Therefore we are able to obtain the power moments, even though the prior density $\rho_{x_{t+1} \mid \mathcal{Y}_{t}}(x)$ doesn't have an analytic form as a function. There have been numerous previous research results using the power moments for Bayesian filtering. For example in the well-known Kalman filter (and its extended forms such as EKF and UKF), approximation of the density is done by a parametric estimation using the first order and second order power moments. However a multivariate Bayesian filter feasible for treating the density of the state 
using higher order moments has not been proposed to the best of our knowledge. In this paper, our goal is to propose a multivariate Bayesian filter which not only is feasible to treat the non-Gaussian state estimation problem, but also has analytic error analyses to measure the performance of filtering.

To perform the density approximation in the following parts of the paper, we first define the equivalence of multivariate density functions in the sense of power moments. 

\begin{definition}
A probability density function, which has the identical $L^{\rho}\left( \kappa \right), \kappa \in \mathcal{J}_{2n}$ as $\rho$, is called an order-$2n$ density surrogate of $\rho$ and denoted by $\rho^{2n}$.
\end{definition}

\begin{algorithm}[t]
	\caption{A general framework for Bayesian filtering with density surrogate at time $t$.}
	\label{alg:algorithm1}
	\KwIn{System parameters: $f_{t}, h_{t}$;\\
	Non-Gaussian densities: $\eta_{t}, \epsilon_{t}$; \\
	Prediction at time $t-1$: $\hat{\rho}_{x_{t} \mid \mathcal{Y}_{t-1}}(x), t > 0, \text{or } \rho_{x_{0}}(x), t = 0 $.}
	\KwOut{Prediction at time $t$: $\hat{\rho}_{x_{t+1} \mid \mathcal{Y}_{t}}(x)$.}  
	\BlankLine
	Step 1: Calculate $\hat{\rho}_{x_{t}|\mathcal{Y}_{t}}$ by (\ref{Update1}) or (\ref{Update2});\\
	Step 2: Calculate $L^{\rho}\left( \kappa \right)$ by (\ref{MomentUpdate}) for $\kappa \in \mathcal{J}_{2n}$;\\
    Step 3: Determine the multivariate order-$2n$ density surrogate $\rho^{2n}_{x_{t+1} \mid \mathcal{Y}_{t}}$, of which the truncated moments are $L^{\rho}\left( \kappa \right), \kappa \in \mathcal{J}_{2n}$. The density estimate at time $t+1$ is then chosen as the density surrogate, i.e., $\hat\rho_{x_{t+1} \mid \mathcal{Y}_{t}} = \rho^{2n}_{x_{t+1} \mid \mathcal{Y}_{t}}$.
\label{Algo1}
\end{algorithm}

By defining the corresponding density estimate as $\hat{\rho}$, we propose a general framework for each iteration of Bayesian filtering with the density surrogate as Algorithm \ref{Algo1}. Now the problem amounts to constructing an order-$2n$ multivariate density surrogate. Since the domain of $\rho$ is $\mathbb{R}^{d}$, the problem becomes a multidimensional Hamburger moment problem \cite{schmudgen2017moment}. In the next section, we will give a formal definition to the multidimensional Hamburger moment problem we will treat, and show the existence of solution to it, i.e., the existence of the multivariate density surrogate.

\section{The Multidimensional Hamburger moment problem and the existence of solution}
\label{TheHamburger}

In this section, we give a formal definition of the Hamburger moment problem to treat and prove the existence of solution to the multidimensional Hamburger moment problem with the power moments. First we give the definition. 

\begin{definition}
A sequence $\bar{\sigma}=\left(\sigma_{\kappa}, \kappa \in \mathcal{J}_{2n}\right)$ is a feasible $2n$ sequence, if there is a random vector $X=\left(X_{1}, \cdots, X_{d}\right)^{\intercal}$ defined on $\mathbb{R}^{d}$ endowed with its Borel sigma algebra of events, whose power moments are given by \eqref{MomentUpdate}, that is,
\begin{equation}
    \sigma_{\kappa} = \sigma_{k_{1}, \cdots, k_{d}}
    =  L^{\rho}\left( \kappa \right) = \int_{\mathbb{R}^{d}} x_{ 1}^{k_{1}} x_{2}^{k_{2}} \cdots x_{d}^{k_{d}} \rho(x) dx
\label{MomentCondition}
\end{equation}
for all $\kappa \in \mathcal{J}_{2n}$. We say that any such random vector $X$ has a $\bar{\sigma}$-feasible distribution and denote this as $X \sim \bar{\sigma}$. 
\label{Def31}
\end{definition}

Next we show the existence of solution to the moment problem defined in Definition \ref{Def31}. 
We note that the true $\rho_{x_{t+1} \mid \mathcal{Y}_{t}}(x)$ is trivially a solution to the moment problem in Definition \ref{Def31}. However we require an analytic function which satisfy the moment constraints. We note that there are numerous methods for functional approximation, however the solutions they provide don't always satisfy the requirement as a state estimation of the Bayesian filter, since there are possibly infinitely many parameters in the solutions, which makes it infeasible to propagate the solutions in the filtering process. Parametrization is then the most significant problem for constructing the density surrogate, which aims to use finitely many parameters to characterize the density. 

Moreover, we are provided with the truncated power moment sequence $\bar{\sigma}$ rather than the full one, which means that there might be infinitely many feasible solutions to this problem. In the following part of this section, we propose to choose proper constraints to parameterize the density surrogate which satisfies the moment conditions. We shall still emphasize here that the parametrization is not unique. Different constraints will yield different parametrizations.

In the next section, we propose to parameterize the multivariate density surrogate, i.e., to derive a unique solution to the moment problem of $\rho_{x_{t+1} \mid \mathcal{Y}_{t}}$.

\section{parametrization of the multivariate density surrogate using power moments}
\label{ParametrizationOf}

In this section, we propose to parameterize the density surrogate, i.e., to derive a unique solution to the multidimensional Hamburger moment problem of $\rho_{x_{t+1} \mid \mathcal{Y}_{t}}$. For simplicity, we omit the subscript $t$ in all the terms of the following part of this section.

Denote the Kronecker product as $\otimes$ and we observe that the moment conditions in \eqref{MomentCondition} can be written in the vector form
\begin{equation}
   \int_{\mathbb{R}^{d}} F(x_{1})\otimes F(x_{2}) \otimes \cdots \otimes F(x_{d}) \rho(x)dx=\breve{\Sigma},
\label{IntegG}
\end{equation}
where 
\begin{equation*}
F(x_{i}) = \begin{bmatrix}
1 & x_{i} & \cdots & x_{i}^{2n-1} & x_{i}^{2n}
\end{bmatrix}^{\intercal}
\end{equation*}
and $\breve{\Sigma}$ is written as
$$
\breve{\Sigma} = \begin{bmatrix}
1 & \sigma_{0,0,\cdots,1} & \cdots & \sigma_{2n,2n,\cdots,2n}
\end{bmatrix}^{\intercal}
$$
with the power moments $\sigma_{k_{1}, \cdots, k_{d}}$ calculated by \eqref{MomentCondition}. We note that the dimension of $\breve{\Sigma}$ is $(2n+1)^{d}$. Then we have that $\breve{\Sigma}$ is in the range of the linear integral operator
$$
\breve{\Gamma}: \rho \mapsto \breve{\Sigma} = \int_{\mathbb{R}^{d}} F(x_{1})\otimes F(x_{2}) \otimes \cdots \otimes F(x_{d}) \rho(x)dx.
$$

Consequently, we have an order $2n$ multidimensional moment problem as defined in Definition \ref{Def31}.

\subsection{A convex optimization scheme}
We note that to directly treat the multivariate Hamburger moment problem of \eqref{IntegG} is not feasible, since there is no prior knowledge of feasible classes of the density. If we assume the density of the state to fall within specific classes of functions, the moment problem then becomes a parametric estimation problem, where the moments are used to estimate the parameters of the parametric models given prior. However in our problem setting, the prior density of the state is only assumed to be Lebesgue integrable with first several orders of power moments being finite. No knowledge of the feasible functions of the density is required. 

Let $\mathcal{P}$ be the space of probability density functions on the Euclidean space $\mathbb{R}^{d}$ with support there, and let $\mathcal{P}_{2n}$ be the subset of all $\rho\in\mathcal{P}$ which have finite power moments at least up to order $2n$ (in addition to $\sigma_{0,\cdots,0}$, which of course is 1). 

Except for the parametric algorithms, density estimation has been done by optimization in previous results \cite{hall1987kullback, li1999mixture, vapnik1999nature}, where Kullback-Leibler distance is used as a distance measure between densities.  Let $\theta$ be an arbitrary density known prior in $\mathcal{P}$ and consider the Kullback-Leibler (KL) distance
\begin{equation}
\label{KL}
\mathbb{KL}(\theta\|\rho)=\int_{\mathbb{R}^{d}} \theta(x) \log \frac{\theta(x)}{\rho(x)} dx
\end{equation}
between $\theta$ and $\rho$. Although not symmetric in its arguments, the KL distance is jointly convex and is widely used in density estimation. In this section, we extend some lines of thoughts of \cite{georgiou2003kullback} to the multivariate case and introduce a parametrization which is induced by the KL distance. However the extension is not at all a trivial problem, of which the details will be given in this section. 

Given the moment constraints \eqref{IntegG}, we first form the Lagrangian
\begin{equation}
L(\rho, \breve{\Lambda})=\mathbb{KL}(\theta \| \rho)+\breve{\Lambda}(\breve{\Gamma}(\rho)-\breve{\Sigma}),
\label{LagrangianOld}
\end{equation}
where
\begin{equation}
\breve{\Lambda} = \begin{bmatrix}
\lambda_{0,0,\cdots,0} & \lambda_{1,0,\cdots,0} & \cdots & \lambda_{2n,2n,\cdots,2n}
\end{bmatrix}
\label{LambdaOri}    
\end{equation}
is the vector-valued Lagrange multipliers, and consider the problem of maximizing the dual functional
\begin{equation}
\label{infL}
\breve{\Lambda} \mapsto \inf _{\rho \in \mathcal{P}_{2n}} L(\rho, \breve{\Lambda}).
\end{equation}

Clearly $\rho\mapsto L(\rho, \breve{\Lambda})$ is strictly convex, so to be able to determine the right member of \eqref{infL}, we must find a $\rho\in\mathcal{P}_{2n}$, for which the directional derivative $\delta L(\rho, \breve{\Lambda} ; \delta\rho) =0$ for all relevant $\delta\rho$. This will further restrict the choice of $\breve{\Lambda}$. By denoting
$$
F(x) = F(x_{1})\otimes F(x_{2}) \otimes \cdots \otimes F(x_{d})
$$
and setting
\begin{equation}
\label{q}
q(x, \breve{\Lambda}):=\breve{\Lambda} F(x),
\end{equation}
we have
$$
L(\rho, \breve{\Lambda})=\int_{\mathbb{R}^{d}} \theta(x)\log \frac{\theta(x)}{\rho(x)}dx +\int_{\mathbb{R}^{d}}q(x, \breve{\Lambda})\rho(x)dx-\breve{\Lambda} \breve{\Sigma},
$$
with the directional derivative
$$
\delta L(\rho, \breve{\Lambda} ; \delta\rho)=\int_{\mathbb{R}^{d}} \delta\rho(x)\left(q(x, \breve{\Lambda})-\frac{\theta(x)}{\rho(x)}\right)dx ,
$$
which has to be zero at a minimum for all variations $\delta\rho$. Clearly this can be achieved only if
\begin{equation}
    \rho(x)=\frac{\theta(x)}{q(x, \breve{\Lambda})}, \quad \forall x\in\mathbb{R}^{d}.
\label{Parametrization}
\end{equation}. 

\subsection{A positive parametrization for the multivariate density}

We have proved that $\rho(x)=\theta(x)/q(x, \breve{\Lambda})$ maximizes the dual functional, however there is still a constraint we need to consider. We note that $\rho(x)$ and $\theta(x)$ are both probability density functions, therefore are both nonnegative on $\mathbb{R}^{d}$. Moreover, $x$ is supported on $\mathbb{R}^{d}$ for the Hamburger moment problem, so then we need to have $\rho(x) > 0, x \in \mathbb{R}^{d}$.

The problem now amounts to characterizing the constraint of $\breve{\Lambda}$, under which the multivariate polynomial $q(x, \breve{\Lambda})$ is positive. However it is a challenging problem of this paper and even of the general multidimensional moment problem. The reason is that useful descriptions of strictly positive polynomials up to a fixed degree $2n$ are missing \cite{schmudgen2017moment}. The positive definiteness of the Hankel matrix is a sufficient condition for the univariate polynomials to be positive, which is used in our previous papers \cite{georgiou2003kullback, wu2023non}. However it is not valid anymore for the multivariate polynomials.

In the multivarate Hamburger moment we consider, the highest order of each variable $x_{i}$ is chosen as $2n$. Indeed, it is a necessary condition for a polynomial $q(x, \breve{\Lambda})$ to be strictly positive everywhere on $\mathbb{R}^{d}$. A polynomial of which the highest order of any variable is odd always has a real zero, and the value of the polynomial changes sign at that point. It is then not feasible to obtain a $q(x, \breve{\Lambda}) > 0$.

With highest order of each variable chosen as $2n$, we propose a new parametrization of $q(x, \breve{\Lambda})$ and give the following theorem which is a strict Positivstellensatz for $q(x, \breve{\Lambda})$. We note that by our choice of $\mathcal{J}_{2n}$, it is feasible for us to write $q(x, \breve{\Lambda})$ in a quadratic form. Denote
\begin{equation*}
G(x_{i}) = \begin{bmatrix}
1 & x_{i} & \cdots & x_{i}^{n-1} & x_{i}^{n}
\end{bmatrix}^{\intercal}
\end{equation*}
and
\begin{equation*}
    G(x) = G(x_{1})\otimes G(x_{2}) \otimes \cdots \otimes G(x_{d}),
\end{equation*}
and we can write
\begin{equation}
    q(x, \breve{\Lambda}) = q(x, \Lambda) = {G^{\intercal}(x)\Lambda G(x)}.
\label{LambdaEquiv}
\end{equation}

Here we note that the dimension of $\breve{\Lambda}$ and that of $\Lambda$ are different, which are $(2n+1)^{d}$ and $(n+1)^{2d}$ respectively. It is obvious that the latter one is always larger than the former one since $n, k \in \mathbb{N}_{0}/\{0\}$. However there are only $(2n+1)^{d}$ moment constraints given by \eqref{MomentCondition}, which means that only $(2n+1)^{d}$ Lagrange multipliers are necessary for the dual functional. For the ease of analyses in the following sections, we write all elements of $\Lambda$ by the elements of $\breve{\Lambda}$.

Denote the $i_{\text{th}}$ element of $G(x)$ as $G(x)_{i}$, and the element of $\Lambda$ at $i_{\text{th}}$ row and $j_{\text{th}}$ column as $\Lambda_{i, j}$. Let the set
$$
\mathcal{R}_{\kappa} := \{\Lambda_{i, j}|G(x)_{i}\cdot G(x)_{j} = x_{ 1}^{k_{1}} x_{2}^{k_{2}} \cdots x_{d}^{k_{d}} \}.
$$

We note that all $\Lambda_{i, j} \in \mathcal{R}_{\kappa}$ corresponds to the same moment constraint

\begin{equation}
    \sigma_{k_{1}, \cdots, k_{d}}
    = \int_{\mathbb{R}^{d}} x_{ 1}^{k_{1}} x_{2}^{k_{2}} \cdots x_{d}^{k_{d}} \rho(x) dx,
\label{MomentConstraint}
\end{equation}
and the cardinality of each $\mathcal{R}_{\kappa}, 0 \leq k_{0}, k_{1}, \cdots, k_{d} \leq 2n$ is $(1+k_{1})(1+k_{2})\cdots(1+k_{d})$. Therefore for each $\Lambda_{i, j} \in \mathcal{R}_{\kappa}$ we have

\begin{equation}
    \Lambda_{i, j} = \frac{\lambda_{k_{1}, k_{2}, \cdots, k_{d}}}{(1+k_{1})(1+k_{2})\cdots(1+k_{d})}.
\label{LambdaBreve}
\end{equation}

By \eqref{LambdaBreve}, we have proposed a one-to-one correspondence between $\breve{\Lambda}$ and $\Lambda$. With the proposed parametrization for $q(x, \Lambda)$, we have the following Positivstellensatz.

\begin{theorem}[Positivstellensatz] $q(x, \Lambda) > 0$ if and only if $\Lambda$ is positive definite.
\label{Thm41}
\end{theorem}

\begin{proof} The necessity is obvious and we need to prove the sufficiency. Denote the eigenvalues of $\Lambda$ as $m_{i}, i \in \mathbb{N}_{0}, i \leq (n+1)^{d}$, and the corresponding eigenvectors as $v_{i}$. We assume that there exists a $q(x, \Lambda) > 0$ of which $\Lambda$ is not positive definite. By eigen decomposition, we can write $\Lambda$ as
\begin{equation*}
    \Lambda = V M V^{\intercal}
\end{equation*}
where
\begin{equation*}
    V := \begin{bmatrix}
    v_{1} & v_{2} & \cdots & v_{(n+1)^{d}}
    \end{bmatrix},
\end{equation*}
and
\begin{equation*}
    M = \begin{bmatrix}
m_{1} & 0 & \cdots & 0\\ 
0 & m_{2} & \cdots & 0\\ 
\vdots & \vdots & \ddots & \\ 
0 & 0 &  & m_{(n+1)^{d}}.
\end{bmatrix}
\end{equation*}

For matrices which are not positive definite, there are at least one nonpositive eigenvalue. We assume $m_{1} \leq \cdots \leq m_{l} \leq 0, l \leq (n+1)^{d}$. And assign $x_{i} = 0, i \in \mathbb{N}_{0}, i \leq (n+1)^{d}$. Then $G(\mathbf{0}) = [1, 0, 0, \cdots, 0]^{\intercal}$. So then we have
\begin{equation}
\begin{aligned}
    & q(\mathbf{0}, \Lambda)\\
    = & G^{\intercal}(\mathbf{0}) V M V^{\intercal} G(\mathbf{0})\\
    = & \operatorname{tr} \left(  V G^{\intercal}(\mathbf{0}) M G(\mathbf{0}) V^{\intercal} \right)\\
    = & G^{\intercal}(\mathbf{0}) M G(\mathbf{0}) \operatorname{tr} \left(  V V^{\intercal} \right).
\end{aligned}
\end{equation}

Since the scalar $G^{\intercal}(\mathbf{0}) M G(\mathbf{0}) \leq 0$ with $m_{1} \leq 0$, and $\operatorname{tr} \left(  V V^{\intercal} \right) > 0$, we have $q(\mathbf{0}, \Lambda) \leq 0$, which contradicts our assumption. Therefore when $\Lambda$ has at least one eigenvalue being nonpositive, there exists at least a point $x = (0, 0, \cdots, 0)$, at which we have $q(x, \Lambda) \leq 0$.
\end{proof}

\begin{remark} 
In the conventional research on the multidimensional truncated moment problem, we care about the existence of the positive measures rather than the parametrization of them. Therefore atomic measures \cite{schmudgen2017moment} (probability mass functions) are always proposed as solutions to the multidimensional truncated  moment problems. There have been few results on parameterizing the densities in a continuous form of function supported on $\mathbb{R}^{d}$. However parameterizing the density and ensuring its positiveness given the power moments are of great significance to the real applications. This theorem ensures the positiveness of $q(x, \Lambda)$ for all $d \in \mathbb{N}_{0}$, which contributes to the multivariate Hamburger moment problem. 
\end{remark}

Moreover, we note that the moment conditions can be written in a matrix representation
\begin{equation}
    \int_{\mathbb{R}^{d}} G(x) \rho(x) G^{\intercal}(x) = \Sigma
\end{equation}
where
\begin{equation*}
    \Sigma_{i, j} = \frac{\sigma_{k_{1}, k_{2}, \cdots, k_{d}}}{(1+k_{1})(1+k_{2})\cdots(1+k_{d})}
\end{equation*}
is the element at the $i_{\text{th}}$ row and the $j_{\text{th}}$ column.

By Section \ref{TheHamburger}, we know that the class of $\rho\in\mathcal{P}$ satisfying \eqref{IntegG} is nonempty. In fact, $\Sigma$ is in the range of the linear integral operator
\begin{equation}
    \Gamma: \rho \mapsto \Sigma = \int_{\mathbb{R}^{d}} G(x) \rho(x) G^{\intercal}(x)dx,
\label{Map}
\end{equation}
which is defined on the space $\mathcal{P}_{2n}$. Since  $\mathcal{P}_{2n}$ is convex, then so is $\operatorname{range}(\Gamma)=\Gamma\mathcal{P}_{2n}$. 

In the following part of this section, we use the notations $\Lambda$ and $\Sigma$ instead of $\breve{\Lambda}$ and $\breve{\Sigma}$. Introducing $\breve{\Lambda}$ and the corresponding $\breve{\Sigma}$ in the previous sections is to emphasize that $(2n+1)^{d}$ is the minimal number of parameters to estimate. We shall always remember that all the elements of $\Lambda$ and $\Sigma$ can be represented by those of $\breve{\Lambda}$ and $\breve{\Sigma}$.

\subsection{Uniqueness of the solution to the convex optimization problem}

Being a multivariate probability density function, the positiveness of the parametrization in \eqref{Parametrization} is ensured by the constraint $\Lambda \succ 0$ as proved in Theorem \ref{Thm41}. We then now state and prove our main results. We emphasize here that we adopt some notations from our previous paper \cite{wu2023non} treating the univariate Bayesian filtering problem, however with different definitions. 

By \eqref{Parametrization} and \eqref{LambdaEquiv}, a possible minimizer must have the form
$$
\rho(x)=\frac{\theta(x)}{q(x)} = \frac{\theta(x)}{G^{\intercal}(x)\Lambda G(x)}.
$$

The Lagrangian in \eqref{LagrangianOld} can then be written as
\begin{equation}
\begin{aligned}
& L(\rho, \Lambda)\\
= & \mathbb{KL}(\theta \| \rho)+ \operatorname{tr}\left (\Lambda(\Gamma(\rho)-\Sigma) \right)\\
= & \int_{\mathbb{R}^{d}} \left(\theta \log \left(G^{\intercal}(x)\Lambda G(x)\right) + \theta\right) dx - \operatorname{tr}\left( \Lambda \Sigma \right).
\label{LagrangianNew}
\end{aligned}
\end{equation}

Then the problem becomes minimizing the dual functional
\begin{equation}
\mathbb{J}_\theta(\Lambda):=\operatorname{tr}(\Lambda \Sigma)-\int_{\mathbb{R}^{d}} \theta(x) \log\left(G^{\intercal}(x)\Lambda G(x)\right) dx.
\label{LossFunc}
\end{equation}

Denote
\begin{equation*}
\label{Lplus}
\mathcal{L}_{+}:=\left\{\Lambda \in \operatorname{range}(\Gamma) \mid \Lambda \succ 0 \right\},
\end{equation*}
and we have the following lemma.

\begin{lemma}
Any stationary point of $\mathbb{J}_{\theta}(\Lambda)$ must satisfy the equation
\begin{equation}
    \omega(\Lambda)=\Sigma,
    \label{Omega}
\end{equation}
where the map $\omega: \mathcal{L}_{+} \rightarrow \mathcal{S}_{+}$ between $\mathcal{L}_{+}$ and $\mathcal{S}_{+}:=\{\Sigma \in \operatorname{range}(\Gamma)\}$ is defined as
$$
\omega: \; \Lambda \mapsto \int_{\mathbb{R}^{d}} G(x) \frac{\theta(x)}{q(x, \Lambda)} G^{\intercal}(x)dx
\label{omega}
$$
with $q$ defined by \eqref{q}. 
\label{Lemma42}
\end{lemma}

\begin{proof}
Using the fact that
\begin{displaymath}
\delta q(\Lambda ; \delta \Lambda)=G^{\intercal} \delta \Lambda G=\operatorname{tr}\left(\delta \Lambda G G^{\intercal}\right)
\end{displaymath}
we have the directional derivative
\begin{equation*}
\delta \mathbb{J}_{\theta}(\Lambda ; \delta \Lambda)=\operatorname{tr}\left(\delta \Lambda\left[\Sigma-\int_{\mathbb{R}^{d}} G(x)\frac{\theta(x)}{q(x, \Lambda)} G^{\intercal}(x)dx\right]\right),
\label{FirstOrder}
\end{equation*}
which is zero for all $\delta \Lambda \in\operatorname{range}(\Gamma)$ if and only if \eqref{Omega} holds. This completes the proof. 
\end{proof}

To prove the existence and uniqueness of the minimum of $\mathbb{J}_{\theta}$, we need to establish that the map $\omega: \mathcal{L}_{+} \rightarrow \mathcal{S}_{+}$ is injective, establishing uniqueness, and surjective, establishing existence. In this way we prove that \eqref{Omega} has a unique solution, and hence that there is a unique minimum of the dual functional $\mathbb{J}_{\theta}$. We start with injectivity.

\begin{lemma}
Suppose $\Lambda \in \text{range}(\Gamma)$. Then the map

\begin{equation}
\label{Lambda2G'LambdaG}
\Lambda \mapsto G^{\intercal} \Lambda G
\end{equation}
is injective.
\label{Lemma43}
\end{lemma}

\begin{proof} 
Since $\Lambda \in \text{range}(\Gamma)$,
$$
\label{xx}
   \Lambda = \int_{\mathbb{R}^{d}} G(y) \psi(y) G^{\intercal}(y)dy
$$
for some $\psi \in \mathcal{P}$.
Suppose $G^{\intercal}\Lambda G = 0$. Then we have $\int_{\mathbb{R}^{d}} G^{\intercal}(x)\Lambda G(x)dx = 0$, and therefore
$$
\begin{aligned}
    & \int_{\mathbb{R}^{d}} G^{\intercal}(x)\Lambda G(x)dx \\
    = & \operatorname{tr} \left( \int_{\mathbb{R}^{d}} G(x)^{\intercal}\int_{\mathbb{R}^{d}} G(y) \psi(y) G(y)^{\intercal}dy\, G(x) dx \right) \\
    = & \int_{\mathbb{R}^{d}} \int_{\mathbb{R}^{d}} [G(x)^TG(y)]^2\psi(y) dxdy =0.
\end{aligned}
$$

Consequently we have $[G(x)^TG(y)]^2\psi(y) = 0$, for all $x,y \in \mathbb{R}^{(n+1)^{d}}$, which clearly implies that $\psi=0$, and hence that $\Lambda = 0$.
Consequently the map \eqref{Lambda2G'LambdaG} is injective, as claimed.
\end{proof}

\begin{theorem}
The map $\omega: \mathcal{L}_{+} \mapsto \mathcal{S}_{+}$ is a homeomorphism. 
\label{Theorem44}
\end{theorem}
\begin{proof}
We first prove that $\omega: \mathcal{L}_{+} \mapsto \mathcal{S}_{+}$ is injective. Suppose that $\omega(\Lambda_1)=\omega(\Lambda_2)$ for some $\Lambda_1$ and $\Lambda_2$ in $\mathcal{L}_+$. We need to show that $\Lambda_1=\Lambda_2$. To this end, note that
$$\Delta \omega = \omega(\Lambda_1)-\omega(\Lambda_2)=\int_{\mathbb{R}^{d}}GG^T\frac{\theta}{q_1q_2}(q_2-q_1)dx=0,$$
where $q_1=G^T\Lambda_1G$ and $q_2=G^T\Lambda_2G$. Considering the element of $\Delta \omega$ at the first row and the first column, we have
\begin{equation}
\Delta \omega _{1, 1} = \int_{\mathbb{R}^{d}}\frac{\theta}{q_1q_2}(q_2-q_1)dx=0.
\label{DeltaOmega}
\end{equation}
Since $\theta$ is a strictly positive probability density function supported on $\mathbb{R}^{d}$, and $q_{1}, q_{2} > 0$, the equality in \eqref{DeltaOmega} is achieved if and only if $q_{1} = q_{2}$. Then by Lemma \ref{Lemma43} this implies that $\Lambda_1=\Lambda_2$, establishing that $\omega$ is injective.

Next, we prove that $\omega: \mathcal{L}_{+} \mapsto \mathcal{S}_{+}$ is surjective. We first note that $\omega$ is continuous and that both sets $\mathcal{L}_{+}$ and $\mathcal{S}_{+}$ are nonempty, convex, and open subsets of the same Euclidean space $\mathbb{R}^{(2n+1)^{d}}$, and hence diffeomorphic to this space. We emphasize again that the dimension of the space is $(2n+1)^{d}$ rather than $(n+1)^{2k}$, since some of the elements in $\Lambda$ and $\Sigma$ are identical. 

For the proof of surjectivity we shall use Corollary 2.3 in \cite{byrnes2007interior}, by which the continuous map $\omega$ is surjective if and only if it is injective and proper, i.e., the inverse image $\omega^{-1}(K)$ is compact for any compact $K$ in $\mathcal{S}_{+}$. See Theorem 2.1 in \cite{byrnes2007interior} for a more general statement.

Now it remains to prove that $\omega$ is proper. To this end, we first note that $\omega^{-1}(K)$ must be bounded, since, as if $\|\Lambda\|\to \infty$,  $\omega(\Lambda)$ would tend to zero, which lies outside $\mathcal{L}_{+}$. Now, consider a Cauchy sequence in $K$, which of course converges to a point in $K$. We need to prove that the inverse image of this sequence is compact. If it is empty or finite, compactness is automatic. So we consider it to be infinite in the following proof. Since $\omega^{-1}(K)$ is bounded, there must be a subsequence $(\lambda_k)$ in $\omega^{-1}(K)$ converging to a point $\lambda\in\mathcal{L}_{+}$. It remains to show that $\lambda\in\omega^{-1}(K)$, i.e., $(\lambda_k)$ does not converge to a boundary point, which here would be $q(x)=0$. However this does not happen since then $\operatorname{det} \omega(\Lambda) \rightarrow \infty$, contradicting boundedness of $\omega^{-1}(K)$. Hence $\omega$ is proper. Therefore, the map $\omega: \mathcal{L}_{+} \rightarrow \mathcal{S}_{+}$ is proved to be homeomorphic.
\end{proof}

Moreover, the dual functional has the following property.
\begin{lemma}
The dual functional $\mathbb{J}_{\theta}(\Lambda)$ is strictly convex.
\label{Lemma45}
\end{lemma}
\begin{proof}
This is equivalent to $\delta^{2} \mathbb{J}_{\theta} > 0$ where
\begin{equation}
\delta^{2} \mathbb{J}_{\theta}(\Lambda; \delta \Lambda)=\int_{\mathbb{R}^{d}} \frac{\theta(x)}{q(x)^{2}}\left(G(x)^{\intercal} \delta\Lambda G(x)\right)^{2}dx.
\label{SecondDeriv}
\end{equation}

By (\ref{SecondDeriv}), we have $\delta^{2} \mathbb{J}_{p} \geq 0$, so it remains to show that
$$
    \delta^{2} \mathbb{J}_{p} > 0, \quad \text{for all $\delta \Lambda \neq \mathbf{0}$},
$$
which follows directly from Lemma \ref{Lemma43}, replacing $\Lambda$ by $\delta\Lambda$.
\end{proof}

By Lemma \ref{Lemma42}, Theorem \ref{Theorem44} and Lemma \ref{Lemma45}, we have the following theorem.

\begin{theorem}
The functional $\mathbb{J}_{\theta}(\Lambda)$ has a unique minimum $\hat{\Lambda} \in\mathcal{L}_{+}$. Moreover
$$
\Gamma\left(\frac{\theta}{G^{\intercal}(x)\hat{\Lambda}G(x)}\right)=\Sigma.
$$
\label{Thm43}
\end{theorem}

By this theorem, 
$$
\hat{\rho}=\frac{\theta}{\hat{q}}, \quad \hat{q}=q(x, \hat{\Lambda})
$$
belongs to $\mathcal{P}_{2n}$ and is a stationary point of $\rho \mapsto L(\rho, \hat{\Lambda})$, which is strictly convex. Consequently
$$
L(\hat{\rho}, \hat{\Lambda}) \leq L(\rho, \hat{\Lambda}), \quad \text { for all } \rho \in \mathcal{P}_{2n}
$$
or, equivalently, since $\Gamma(\hat{\rho})=\Sigma$,

$$
\mathbb{KL}(\theta\| \hat{\rho}) \leq \mathbb{KL}(\theta \| \rho)
$$
for all $\rho\in \mathcal{P}_{2n}$ satisfying the constraint $\Gamma(\rho)=\Sigma$. The above holds with equality if and only if $\rho=\hat{\rho}$. Finally, a complete solution is given in the following theorem for density parametrization using the multivariate power moments.

\begin{theorem}
Let $\Gamma$ be defined by (\ref{Map}) and $\Lambda$ defined by \eqref{LambdaBreve}. Given any $\theta \in \mathcal{P}$ and any $\Sigma$ with $\sigma_{\kappa}$ calculated by \eqref{MomentUpdate}, there is a unique $\rho \in \mathcal{P}_{2n}$ that minimizes \eqref{KL}
subject to $\Gamma(\rho)=\Sigma$, i.e., subject to \eqref{IntegG}, namely
\begin{equation}
    \hat{\rho}=\frac{\theta}{q(x, \hat{\Lambda})},
\label{MiniForm}
\end{equation}
where $\hat{\Lambda}$ is the unique solution to the problem of minimizing $\mathbb{J}_{\theta}$ in \eqref{LossFunc} over all $\Lambda \in \mathcal{L}_{+}$. 
\label{Thm47}
\end{theorem}

Consequently, the dual problem provides us with an approach to compute the unique $\hat{\rho}$ that minimizes the Kullback-Leibler distance $\mathbb{KL}(\theta \| \rho)$ subject to the constraint $\Gamma(\rho)=\Sigma$. 

A parametrization using power moments for the density surrogate has been proposed in Theorem \ref{Thm47}. Apply the proposed parametrization of the density surrogate to estimate $\rho_{x_{t+1} \mid \mathcal{Y}_{t}}(x)$ , we are able to design Algorithm 2 following the framework of Algorithm 1. However, the proposed filter is not the only feasible Bayesian filter by the multivariate power moments. By proposing different density parametrization strategies using power moments as Step 3 in Algorithm 1, more of such type of filters can be developed.

\begin{algorithm}[t]
	\caption{Bayesian filtering with density surrogate using power moments at time $t$.}
	\label{alg:algorithm2}
	\KwIn{System parameters: $f_{t}, h_{t}$;\\
	Non-Gaussian densities: $\eta_{t}, \epsilon_{t}$; \\
	Prediction at time $t-1$: $\hat{\rho}_{x_{t} \mid \mathcal{Y}_{t-1}}(x), t > 0, \text{or } \rho_{x_{0}}(x), t = 0 $.}
	\KwOut{Prediction at time $t$: $\hat{\rho}_{x_{t+1} \mid \mathcal{Y}_{t}}(x)$.}  
	\BlankLine
	Step 1: Calculate $\hat{\rho}_{x_{t}|\mathcal{Y}_{t}}$ by (\ref{Update1}) or (\ref{Update2});\\
	Step 2: Calculate $\sigma_{\kappa}, \kappa \in \mathcal{J}_{2n}$ of $\int_{\mathbb{R}^{d}} \hat \rho_{x_{t} \mid \mathcal{Y}_{t}}\left(f^{-1}_{t}(\varepsilon)\right) \rho_{\eta_{t}}(x-\varepsilon) d\varepsilon$ by (\ref{MomentUpdate});\\
    Step 3: Do the optimization and obtain $\hat{\Lambda}$ which minimizes \eqref{LossFunc}. $\hat{\rho}_{x_{t+1} \mid \mathcal{Y}_{t}}(x) = \theta(x)/(G^{\intercal}(x)\hat{\Lambda} G(x))$ is the order-$2n$ density surrogate of $\int_{\mathbb{R}^{d}} \hat \rho_{x_{t} \mid \mathcal{Y}_{t}}\left(f^{-1}_{t}(\varepsilon)\right) \rho_{\eta_{t}}(x-\varepsilon) d\varepsilon$.
\label{Algo2}
\end{algorithm}

\section{Error analyses of the density surrogate}
\label{MomentEstimation}

In this section, we will first analyze the error propagation of the power moments using the density surrogate. It has been proved in \cite{wu2023non} that all power moments of $\hat{x}$ exist and are finite, i.e. $\hat{\rho}(x)$ is light-tailed, if and only if $\theta$ is a sub-Gaussian distribution. Sub-Gaussian refers to that the tails of a distribution are dominated by those of a Gaussian distribution, i.e., decay at least as fast as a Gaussian. We will first analyze the error propagation of $\hat{x}$ of which all the power moments exist and are finite. We note that most Bayesian filters satisfy this condition, such as the well-known Kalman filter and the particle filter.

\begin{theorem}
Assume that all power moments of the true density $\rho_{x_{t+1}|\mathcal{Y}_t}$ and the corresponding estimate $\hat{\rho}_{x_{t+1}|\mathcal{Y}_t}$ exist and are finite. Suppose $\hat{\rho}_{x_1|\mathcal{Y}_0}$ to be a surrogate for $\rho_{x_1|\mathcal{Y}_0}$, and let $\hat{\rho}_{x_t|\mathcal{Y}_t}, \hat{\rho}_{x_{t+1}|\mathcal{Y}_t}$ be obtained from Algorithm 1 for $t = 2, 3, \cdots$. Then the power moments of $\hat{\rho}_{x_t|\mathcal{Y}_t}$ and $\hat{\rho}_{x_{t+1}|\mathcal{Y}_t}$ up to order $2n$ are asymptotically unbiased and approximately those of the density surrogates $\rho_{x_t|\mathcal{Y}_t}$ and $\rho_{x_{t+1}|\mathcal{Y}_t}$ respectively with a large enough $n$.
\label{Thm51}
\end{theorem}

\begin{proof} 
For the sake of simplicity, we omit the normalizing factor in the measurement update equations \eqref{Update1} and \eqref{Update2}. It is straightforward to verify that it has no effect on the following results in this section. The first $2n$ moment terms of $\rho_{x_{1} \mid \mathcal{Y}_{0}}$ are identical to $\hat \rho_{x_{1} \mid \mathcal{Y}_{0}}$ after the first time update, i.e.,
\begin{equation}
    L^{\rho_{1|0}}\left( \kappa \right) = L^{\hat{\rho}_{1|0}}\left( \kappa \right), \quad \kappa \in \mathcal{J}_{2n}.
\label{MomentEqual}
\end{equation}

For each $\mu \in \mathcal{J}_{2n}$, we can write the moment terms of $\rho_{x_{1} \mid \mathcal{Y}_{1}}$ as
\begin{equation*}
    L^{\rho_{1|1}}\left( \kappa \right)
    = \int_{\mathbb{R}^{d}} x_{ 1}^{k_{1}} x_{2}^{k_{2}} \cdots x_{d}^{k_{d}} \rho_{\epsilon_{1}}\left(y_{1}-f_{1} (x)\right) \rho_{x_{1} \mid \mathcal{Y}_{0}}(x)dx
\end{equation*}
for all $\kappa \in \mathcal{J}_{2n}$. And those of $\hat \rho_{x_{1} \mid \mathcal{Y}_{1}}$ as,
\begin{equation*}
    L^{\hat{\rho}_{1|1}}\left( \kappa \right)
    = \int_{\mathbb{R}^{d}} x_{ 1}^{k_{1}} x_{2}^{k_{2}} \cdots x_{d}^{k_{d}} \rho_{\epsilon_{1}}\left(y_{1}-f_{1} (x)\right) \hat{\rho}_{x_{1} \mid \mathcal{Y}_{0}}(x)dx
\end{equation*}
for all $\kappa \in \mathcal{J}_{2n}$. Therefore we have,
\begin{equation}
\begin{aligned}
    & L^{\rho_{1|1}}\left( \kappa \right) - L^{\hat{\rho}_{1|1}}\left( \kappa \right)\\
    = & \int_{\mathbb{R}^{d}} x_{ 1}^{k_{1}}  \cdots x_{d}^{k_{d}} \rho_{\epsilon_{1}}\left(y_{1}-f_{1} (x)\right) \left ( \rho_{x_{1} \mid \mathcal{Y}_{0}} - \hat \rho_{x_{1} \mid \mathcal{Y}_{0}} \right )dx.
\end{aligned}
\label{Lminus}
\end{equation}

It is not feasible for us to calculate it directly, since we only have the moment constraints \eqref{MomentEqual}. We naturally consider decomposing the $\rho_{\epsilon_{1}}\left(y_{1}-f_{1} (x)\right)$ into a polynomial, so then \eqref{Lminus} is able to be written in the form of a weighted sum of power moments of $ \rho_{x_{1} \mid \mathcal{Y}_{0}} - \hat \rho_{x_{1} \mid \mathcal{Y}_{0}}$. By generalizing Exercise 13.12 in \cite{rudin_2015} to the multivariate case, it is feasible for us to write the Lebesgue integrable function $\rho_{\epsilon_{1}}\left(y_{1}-h_{1} x\right)$ as a polynomial of multiple variables, which is denoted as $\tilde{\rho}_{\epsilon_{1}}\left(y_{1}-h_{1} (x)\right)$. By Taylor expansion, we write it as
$$
\begin{aligned}
    & \tilde{\rho}_{\epsilon_{1}}\left(y_{1}-f_{1} (x)\right)\\ = & \tilde{\rho}_{\varepsilon_{1}}(y_{1})+\left(\partial_{l_{1}} \tilde{\rho}_{\epsilon_{1}}\right)(y_{1}) x_{i}+\frac{1}{2 !}\left(\partial_{l_{1} l_{2}} \tilde{\rho}_{\epsilon_{1}}\right)(y_{1}) x_{l_{1}} x_{l_{2}}\\
    + & \frac{1}{3 !}\left(\partial_{l_{1} l_{2} l_{3}} \tilde{\rho}_{\epsilon_{1}}\right)(y_{1}) x_{l_{1}} x_{l_{2}} x_{l_{3}}+\cdots\\
    = & \tilde{\rho}_{\epsilon_{1}}(y_{1}) + \sum_{i = 1}^{+\infty} \frac{1}{i!}\left(\partial_{l_{1:i}} \tilde{\rho}_{\epsilon_{1}}\right)(y_{1}) x_{l_{1:i}}
\end{aligned}
$$
where
$$
    \left(\partial_{l_{1:i}} \tilde{\rho}_{\epsilon_{1}}\right)(y_{1}) x_{l_{1:i}} := \left(\partial_{l_{1} l_{2} \cdots l_{i}} \tilde{\rho}_{\epsilon_{1}}\right)(y_{1}) x_{l_{1}} x_{l_{2}} \cdots x_{l_{i}}.
$$

We note that
\begin{equation*}
\begin{aligned}
    & \mathbb{E}\left ( x_{1}^{\kappa}|\mathcal{Y}_{1} \right ) - \mathbb{E}\left ( \hat{x}_{1}^{\kappa}|\mathcal{Y}_{1} \right ) \\
    & = \sum_{i = 1}^{+\infty} \frac{\left(\partial_{l_{1:i}} \tilde{\rho}_{\epsilon_{1}}\right)(y_{1})}{i!} \int_{\mathbb{R}^{d}} x_{l_{1:k+i}} \left ( \rho_{x_{1} \mid \mathcal{Y}_{0}} - \hat \rho_{x_{1} \mid \mathcal{Y}_{0}} \right )dx \\
    & + \int_{\mathbb{R}^{d}} \left ( \rho_{x_{1} \mid \mathcal{Y}_{0}} - \hat \rho_{x_{1} \mid \mathcal{Y}_{0}} \right )dx \\
    & = \sum_{i = (2n)^{d} + 1}^{+\infty} \frac{\left(\partial_{l_{1:i}} \tilde{\rho}_{\epsilon_{1}}\right)(y_{1})}{(i-k)!} \int_{\mathbb{R}^{d}} x_{l_{1:i}} \left ( \rho_{x_{1} \mid \mathcal{Y}_{0}} - \hat \rho_{x_{1} \mid \mathcal{Y}_{0}} \right )dx,
\end{aligned}
\end{equation*}
for $k = \sum_{i = 1}^{d} k_{i} \leq (2n)^{d}$. Therefore we obtain
\begin{equation*}
   \mathbb{E}\left ( x_{1}^{\kappa}|\mathcal{Y}_{1} \right ) = \lim_{n \rightarrow +\infty} \mathbb{E}\left ( \hat{x}_{1}^{\kappa}|\mathcal{Y}_{1} \right ).
\end{equation*}

We also note that $\int_{\mathbb{R}^{d}} x_{ 1}^{k_{1}} x_{2}^{k_{2}} \cdots x_{d}^{k_{d}} \rho_{\eta_{t}}(x-\varepsilon) dx$ in \eqref{MomentUpdate} is indeed a polynomial of $\varepsilon_{i}, i = 1, \cdots, d$, of which the highest order of $\varepsilon_{i}$ in each term is $k_{i} \leq 2n$, $i = 1, \cdots, d$. Therefore by (\ref{MomentUpdate}) we obtain
\begin{equation*}
    \mathbb{E}\left ( x_{2}^{\kappa}|\mathcal{Y}_{1} \right ) = \lim_{n \rightarrow +\infty}\mathbb{E}\left ( \hat{x}_{2}^{\kappa}|\mathcal{Y}_{1} \right ).
\end{equation*}

Similarly we have
\begin{equation*}
    \mathbb{E}\left ( x_{t}^{\kappa}|\mathcal{Y}_{t} \right ) = \lim_{n \rightarrow +\infty}\mathbb{E}\left ( \hat{x}_{t}^{\kappa}|\mathcal{Y}_{t} \right ),
\end{equation*}
and
\begin{equation*}
    \mathbb{E}\left ( x_{t+1}^{\kappa}|\mathcal{Y}_{t} \right ) = \lim_{n \rightarrow +\infty}\mathbb{E}\left ( \hat{x}_{t+1}^{\kappa}|\mathcal{Y}_{t} \right ),
\end{equation*}
which proves the asymptotic unbiasedness of the moment estimates throughout the filtering process as $n \rightarrow \infty$. By properly selecting a large enough $n$, we have
\begin{equation*}
    \mathbb{E}\left ( x_{t}^{\kappa}|\mathcal{Y}_{t} \right ) \approx \mathbb{E}\left ( \hat{x}_{t}^{\kappa}|\mathcal{Y}_{t} \right ),\ k_{i} \leq 2n, i=1, \cdots, d,
\end{equation*}
and
\begin{equation*}
    \mathbb{E}\left ( x_{t+1}^{\kappa}|\mathcal{Y}_{t} \right ) \approx \mathbb{E}\left ( \hat{x}_{t+1}^{\kappa}|\mathcal{Y}_{t} \right ),\ k_{i} \leq 2n, i=1, \cdots, d.
\end{equation*}
\end{proof}

Theorem \ref{Thm51} proves that the moment terms up to order $2n$ of the estimated prior densities with the density surrogate are asymptotically unbiased and approximately identical to the true ones throughout the whole filtering process. It reveals the fact that with a sub-Gaussian $\theta$, approximation using the truncated power moments doesn't introduce significant cumulative errors to the moment terms up to order $2n$ of the estimated pdfs, with a proper choice of $n$. 

Now we consider $\hat{\rho}$ to be heavy-tailed, which is the case for some scenarios such as financial engineering. Since the power moments of $\hat{\rho}$ are not all finite, i.e.,
$$
\int_{\mathbb{R}^{d}} x^{\kappa} \left ( \rho_{x_{t+1} \mid \mathcal{Y}_{t}} - \hat \rho_{x_{t+1} \mid \mathcal{Y}_{t}} \right )dx = \infty, \exists k_{i} \in \mathbb{N}_{0},
$$
it is not feasible for us to apply Taylor expansion to analyze the error propagation. However, we note that
\begin{equation}
\begin{aligned}
    & \left|\mathbb{E}\left ( x_{t+1}^{\kappa}|\mathcal{Y}_{t+1} \right ) - \mathbb{E}\left ( \hat{x}_{t+1}^{\kappa}|\mathcal{Y}_{t+1} \right )\right|\\
    & = \left|\int_{\mathbb{R}^{d}} x^{\kappa} \rho_{\epsilon_{t+1}}\left(y_{t+1}-h_{t+1} x\right) \left ( \rho_{x_{t+1} \mid \mathcal{Y}_{t}} - \hat \rho_{x_{t+1} \mid \mathcal{Y}_{t}} \right )dx\right|\\
    & \leq \int_{\mathbb{R}^{d}} \left|x^{\kappa} \rho_{\epsilon_{t+1}}\left(y_{t+1}-h_{t+1} x\right) \left ( \rho_{x_{t+1} \mid \mathcal{Y}_{t}} - \hat \rho_{x_{t+1} \mid \mathcal{Y}_{t}} \right )\right|dx\\
    & = \int_{\mathbb{R}^{d}} \left| x \right|^{\kappa}\rho_{\epsilon_{t+1}}\left(y_{t+1}-h_{t+1} x\right)\left|  \rho_{x_{t+1} \mid \mathcal{Y}_{t}} - \hat \rho_{x_{t+1} \mid \mathcal{Y}_{t}} \right|dx.
\end{aligned}
\end{equation}

Since $C_{k}:=\int_{\mathbb{R}}|x|^{\kappa} \rho_{\epsilon_{t+1}}\left(y_{t+1}-h_{t+1} x\right) d x$ is a constant unrelated to $\hat{\rho}_{x_{t+1} \mid \mathcal{Y}_{t}}$, we have
$$
\begin{array}{l}
\left|\mathbb{E}\left(x_{t+1}^{\kappa} \mid \mathcal{Y}_{t+1}\right)-\mathbb{E}\left(\hat{x}_{t+1}^{\kappa} \mid \mathcal{Y}_{t+1}\right)\right| \\
\leq C_{k} \max _{x}\left|\rho_{x_{t+1} \mid \mathcal{Y}_{t}}-\hat{\rho}_{x_{t+1} \mid \mathcal{Y}_{t}}\right|.
\end{array}
$$

Consequently, we have proven the following theorem.

\begin{theorem}
The errors of the power moments of $\hat{\rho}_{x_{t+1} \mid \mathcal{Y}_{t+1}}$ are each bounded by a value which is proportional to the $L_{\infty}$ norm of the error of the density surrogate $\hat{\rho}_{x_{t+1} \mid \mathcal{Y}_{t}}$.
\end{theorem}

It reveals the fact that with a relatively small L-infinity norm of $\rho_{x_{1} \mid \mathcal{Y}_{0}} - \hat \rho_{x_{1} \mid \mathcal{Y}_{0}}$, the error of moment estimation will be tolerable.

In the previous sections, a parametrization using power moments for the multivariate density surrogate and a corresponding multivariate Bayesian filter have been proposed. The proposed Bayesian filter uses a continuous form of function to characterize the density of the state, which has not been proposed in the previous results with no constraints on feasible density functions. The continuous parametrization of the multivariate density function also makes it feasible to analyze the error of density estimate
quantitatively.  

To the best of our knowledge, an error upper bound for the multivariate state estimate has not been established in stochastic filtering. The reason is that a continuous form of parametrization of the system state has not been proposed.  In this section, we propose an error upper bound of $\hat{\rho}(x)$ in the sense of total variation distance, which is a measure widely used in the moment problem \cite{Aldo2003A, tagliani2003maximum}. This upper bound distinguishes our proposed filter from other multivariate Bayesian filters.

The total variation distance between the density estimate $\hat{\rho}$ and the true density $\rho$ is defined as follows:
\begin{equation}
    V(\hat\rho, \rho) = \sup_{x} \left|\int_{\left(-\infty, x \right] }(\hat \rho - \rho) d x\right|
    = \sup_{x} \left| F_{\hat \rho} - F_{\rho} \right|
\label{lim}    
\end{equation}
where $F_{\hat \rho}$ and $F_{\rho}$ are the two distribution functions of $\hat \rho$ and $\rho$.

In \cite{Aldo2003A}, Shannon-entropy is used to calculate the upper bound of the total variation distance. The Shannon-entropy \cite{shannon1948mathematical} is defined as

$$
    H[\rho] = - \int_{\mathbb{R}^{d}}\rho(x) \log \rho(x)dx.
$$

Unlike the univariate case, $\rho(x)$ is a multivariate distribution in this paper. To the best of our knowledge, there has not been an existing result of a Shannon-entropy maximizing distribution $\breve{\rho}$ as for the multivariate case, However with the following theorem, it is feasible for us to obtain a Shannon-entropy maximizing distribution for the multivariate problem.

\begin{theorem}[Full chain rule \cite{polyanskiy2014lecture}]
$$
H\left(\rho(x_{1:d})\right)=\sum_{i=1}^{d} H\left(\rho(x_{i}| x_{1:i-1}) \right) \leq \sum_{i=1}^{d} H\left(\rho(x_{i})\right)
$$
with equality iff $x_{1}, \cdots, x_{n}$ are mutually independent.
\label{Thm53}
\end{theorem}

By Theorem \ref{Thm53}, it is feasible for us to turn the entropy maximizing problem of the joint distribution into maximizing the entropy of the marginal distribution of each $x_{i}, i = 1, \cdots, d$, and each $x_{i}$ are independent from each other. So then the joint Shannon-entropy maximizing distribution $\breve{\rho}$ can be written as
$$
    \breve{\rho}(x_{1:d}) = \arg\max_{\rho} H \left(\rho(x_{1:d}) \right) = \prod_{i=1}^{d} \breve{\rho}(x_{i})
$$
where
$$
    \breve{\rho}(x_{i}) = \arg\max_{\rho}H\left(\rho(x_{i})\right).
$$

The univariate Shannon-entropy maximizing distribution $\breve{\rho}$, of which the moments are calculated by \eqref{MomentUpdate}, has the following density function \cite{kapur1992entropy},
$$
    \breve{\rho}(x_{i}) = \exp \left ( - \sum_{j = 0}^{2n} \lambda_{i, j} x_{i}^{j} \right )
$$
where $\lambda_{i, 0}, \cdots, \lambda_{i, 2n}$ are determined by the constraints
\begin{equation}
    \int_{\mathbb{R}^{d}} x_{i}^{l} \exp \left ( - \sum_{j = 0}^{2n} \lambda_{i, j} x_{i}^{j} \right )dx_{i}=\hat{\sigma}_{\kappa_{i, l}},
\label{sigmakappa}
\end{equation}
and $\kappa_{i, l} := 
 \{ \left(k_{1}, \cdots, k_{n}\right) \mid k_{i} = l \leq 2n, k_{j} = 0, j \neq i \}
$
for $i \leq d$.

Therefore we have

\begin{equation}
\begin{aligned}
  & H(\breve{\rho})\\
  = & - \int_{\mathbb{R}^{d}} \prod_{i = 1}^{d} \breve{\rho}(x_{i}) \log \left( \prod_{i = 1}^{d} \breve{\rho}(x_{i}) \right) dx_{1:d}\\
  = & - \int_{\mathbb{R}^{d}} \prod_{i = 1}^{d} \breve{\rho}(x_{i}) \sum_{i = 1}^{d} \log \breve{\rho}(x_{i})dx_{1:d}\\
  = & - \sum_{i = 0}^{d} \int_{\mathbb{R}^{d}} \breve{\rho}(x_{i}) \log \breve{\rho}(x_{i}) dx_{i}\\
  = & - \sum_{i = 0}^{d} \int_{\mathbb{R}^{d}} \breve{\rho}(x_{i}) \cdot \left( - \sum_{j = 0}^{2n} \lambda_{i, j} x_{i}^{j} \right) dx_{i}\\
  = & \sum_{i = 0}^{d}\sum_{l = 0}^{2n} \lambda_{i, l} \sigma_{\kappa_{i, l}}   
\end{aligned}
\label{Hbreverho}
\end{equation}
where the fifth equation is by \eqref{sigmakappa}. Referring to \cite{Aldo2003A}, the KL distance between the true density and the Shannon-entropy maximizing density can be written as

$$
\begin{aligned}
    & \mathbb{KL} \left(\rho \| \breve{\rho}\right)\\
    = & \int_{\mathbb{R}^{d}} \rho(x) \log \frac{\rho(x)}{\breve{\rho}(x)} d x \\
    = & \int_{\mathbb{R}^{d}} \rho(x) \log \rho(x)dx - \int_{\mathbb{R}^{d}} \rho(x) \log \breve{\rho}(x)dx \\
    = & - H\left [ \rho \right ] + \sum_{i = 0}^{d}\sum_{l = 0}^{2n} \lambda_{i, l} \sigma_{\kappa_{i, l}}.
\end{aligned}
$$

However, if $\theta$ is a sub-Gaussian and $n$ is sufficiently large, $\hat{\sigma}_{\kappa_{i, l}}$ is approximately equal to $\sigma_{\kappa_{i, l}}$ for $i=0,1, \ldots 2 n\left(\sigma_{\kappa_{i, l}} \approx \hat{\sigma}_{\kappa_{i, l}}\right)$ by Theorem \ref{Thm51}, and hence
$$
\mathbb{K} \mathbb{L}(\rho \| \breve{\rho}) \approx H[\breve{\rho}]-H[\rho] .
$$
Similarly, we obtain
$$
\mathbb{K} \mathbb{L}(\hat{\rho} \| \breve{\rho}) \approx H[\breve{\rho}]-H[\hat{\rho}] .
$$

Further by \cite{1970Correction, Aldo2003A}, we have
$$
\begin{aligned}
    & V \left ( \breve{\rho}, \hat{\rho}\right )\\
    \leq & 3\left[-1+\left\{1+\frac{4}{9} \mathbb{KL} \left(\hat{\rho} \| \breve{\rho} \right)\right\}^{1 / 2}\right]^{1 / 2} \\
    = & 3\left[-1+\left\{1+\frac{4}{9} \left ( H\left [ \breve{\rho} \right ] - H\left [ \hat{\rho} \right ] \right )\right\}^{1 / 2}\right]^{1 / 2}
\label{Vbound}
\end{aligned}
$$
and
$$
\begin{aligned}
    & V \left ( \breve{\rho}, \rho \right )\\
    \leq & 3\left[-1+\left\{1+\frac{4}{9} \mathbb{KL} \left(\rho \| \breve{\rho} \right)\right\}^{1 / 2}\right]^{1 / 2} \\
    = & 3\left[-1+\left\{1+\frac{4}{9} \left ( H\left [ \breve{\rho} \right ] - H\left [ \rho \right ] \right )\right\}^{1 / 2}\right]^{1 / 2}.
\label{Vbound2}
\end{aligned}
$$

Then we obtain the  upper bound of the error
$$
\begin{aligned}
& V \left ( \hat{\rho}, \rho \right ) \\
= & \sup_{x}|F_{\hat{\rho}}\left ( x \right )-F_{\rho}\left ( x \right )| \\
\leq & \sup_{x}\left(\left|F_{\hat{\rho}}\left ( x \right )-F_{\breve{\rho}}\left ( x \right )\right|+\left|F_{\breve{\rho}}(x)-F_{\rho(x)}\right|\right) \\
\leq & \sup_{x} \left|F_{\hat{\rho}}\left ( x \right )-F_{\breve{\rho}}\left ( x \right )\right|+\sup_{x} \left|F_{\breve{\rho}}(x)-F_{\rho(x)}\right| \\
= & V \left ( \breve{\rho}, \hat{\rho} \right ) + V \left ( \breve{\rho}, \rho \right ) \\
\leq & 3 \left[-1+\left\{1+\frac{4}{9}\left(H\left [ \breve{\rho} \right ] - H\left [ \hat{\rho} \right ]\right)\right\}^{1 / 2}\right]^{1 / 2} \\
+ & 3\left[-1+\left\{1+\frac{4}{9}\left(H\left [ \breve{\rho} \right ] - H\left [ \rho \right ]\right)\right\}^{1 / 2}\right]^{1 / 2}.
\end{aligned}
\label{UpperBoundUnbiased}
$$

In conclusion, we have proposed the error of the moment estimates given that the prior is either sub-Gaussian or not. And we have put forward a error upper bound of the density estimate. To our knowledge, such an error upper bound has not been proposed for the multivariate Bayesian filtering, without assuming the density function to fall within specific classes, in the previous results. 

\section{Continuous vs discrete: a discussion}

In the previous sections, a novel non-Gaussian Bayesian filter is proposed, of which the system state is parameterized as a continuous function. However the detailed treatments in the previous sections may have concealed the core idea of the proposed filter. In this section, we compare the proposed Bayesian filter to the existing results in a more conceptual manner, which aims to provide a bigger picture of the research on the non-Gaussian filtering and emphasize the significance of our proposed Bayesian filter.

The problem we treat doesn't restrict the non-Gaussian density to fall within specific classes of function (in our setting it is only assumed to be Lebesgue integrable of which first several orders of power moments exist and are finite), estimating the intractable prior density in the time update step is indeed an infinite-dimensional problem. The particle filter treats this estimation problem using discrete points without any assumption on the form of function of the prior density at each time step, which also turns the infinite-dimensional problem into a finite dimensional and tractable one. However characterizing the densities by discrete points requires massive particles to store the probability values of the states. The problem is even worse with the increase of dimensions, which is due to the curse of dimensionality.

Fourier decomposition was developed to map the data samples to the frequency domain, which provides an equivalence between the data samples and the spectral density. By the form of a linear integral operator, it is able to characterize the global property of data samples by a limited number of Fourier coefficients. The density estimation is then finite-dimensional and tractable. As to overcome the disadvantage of the discrete methods, there have been several previous endeavors trying to use limited number of terms in the frequency domain to characterize a wider class of distributions in stochastic filtering. For example, a univariate Bayesian filter was proposed using a state-space calculus scheme to treat the filtering problem in the frequency domain \cite{Aldo2003A}. However the dimension of the state-space in each step keeps increasing which makes it infeasible for the task of filtering. Even a dimension reduction scheme was proposed (a prototype algorithm which might be problematic in real applications), it is difficult to quantitatively analyze the error of estimation introduced by dimension reduction (unlike other density estimation problems, errors in the previous steps will have cumulative effects on the following steps for stochastic filtering).

The power moments are used to characterize the intractable prior density $\rho_{x_{t+1} \mid \mathcal{Y}_{t}}$ in this paper. The power moments have the form of linear integral operators similar to the Fourier coefficients, which is then able to characterize the global property of the density to be estimated. However unlike the Fourier coefficients, the basis functions of the power moments are not orthogonal to each other and their norms are not $1$, which makes it more complicated to use them for density estimation. With the proposed algorithm, it is feasible for us to treat the multivariate density estimation problem using the power moments with arbitrary number of variables. The existence, Positivstellensatz and uniqueness of the solution are all proved. These proofs serve as solid foundations of Bayesian filtering using power moments. 

In conclusion, the particle filter and our proposed filter represent two approaches to treat the infinite-dimensional density estimation problem. Since the estimation problem is intrinsically infinite-dimensional, neither of them are optimal filters. However the prior density estimates of them are both asymptotically convergent to the true density, with the number of particles or the moment terms used tending to infinity. Characterizing the intractable prior density by discrete points or power moments are both able to turn the problem into a finite-dimensional and tractable one. However, most existing results on non-Gaussian Bayesian filtering are focused on the discrete representation of the density. With the results of this paper, we would like to see more attention drawn to parameterizing the density in a continuous form of function for Bayesian filtering.

\section{Simulation details and results}

In the previous sections, a Bayesian filter with the density parameterized by using the power moments has been proposed. However, there are still several details to note when implementing the filter. This will be done in this section, where we will provide simulation results to validate the filter we propose.

The first problem is the choice of $\theta(x)$. Mathematically speaking, the choice is arbitrary by Theorem \ref{Thm47}. However, as to achieve a faster convergence rate for optimization in practice, $\theta$ can usually be chosen as a multivariate Gaussian density function, of which the variables are independent from each other, i.e., $\theta(x)= \prod_{i=1}^{d} \theta(x_{i})$. It ensures the existence the finite power moments of $\hat{\rho}(x)$ up to order $2n$. Therefore, the problem reduces to determining the mean and variance of each marginal distribution $\theta(x_{i}), i = 1, \cdots, d$.

The first and second order power moments, i.e., $\sigma_{\kappa_{i, 1}}, \sigma_{\kappa_{i, 2}}$ of the density to be estimated can be calculated by (5). In practice, we can choose $m_{i} = \sigma_{\kappa_{i, 1}}$ and $\sigma_{i}^{2}>\sigma_{\kappa_{i, 2}}$ and determine $\theta(x_{i})=\mathcal{N}\left(m_{i}, \sigma_{i}^{2}\right)$. Here we note that a relatively large variance $\sigma_{i}^{2}$ is to better adjust to the densities with multiple modes.

In the following parts of this section, we will perform two types of experiments to validate the performance of our proposed filtering scheme. The core idea of the paper is to parameterize the intractable prior density by multivariate power moments up to an order. We will first simulate it to see the performance of the algorithm with different choice of the order. Then we will provide the simulation results of the proposed filter in a robot localization problem, which are compared to other prevailing methods.

\subsection{Density estimation by multivariate power moments}

In this simulation, we first estimate the prior density by multivariate power moments. We note that the prior density $\rho_{x_{t+1} \mid \mathcal{Y}_{t}}(x)$ doesn't always have an explicit form of function, i.e., it is not always feasible for us to obtain the true system states. It makes comparing the estimates of the prior density to the true ones infeasible. However, we note that when $\eta_{t}$ is a discrete random variable,  the prior density $\rho_{x_{t+1} \mid \mathcal{Y}_{t}}(x)$ can be written as
\begin{equation}
    \rho_{x_{t+1} \mid \mathcal{Y}_{t}}(x) = \sum_{i=1}^{m} \rho _{i} \cdot \rho_{\eta_{t}}(x-\varepsilon_{i})
\end{equation}
which is a mixture of densities and has an analytic form as a function. In order to compare the density estimates to the true density for validating the performance of the proposed surrogates, we simulate the mixture of densities in the following part of this section. For the ease of visualization, the state $x$ of the examples in this section are all chosen as two dimensional. However we note that our algorithm can treat the filtering problem with system state of arbitrary dimensions. 

In the first two examples, we simulate a mixture of Gaussians. We emphasize here that our proposed density surrogate doesn't require any prior knowledge, such as how many Gaussians there are in the prior density $\rho_{x_{t+1} \mid \mathcal{Y}_{t}}$ to be estimated, or the types of functions in the mixture of densities, which distinguishes the algorithm from other existing estimation methods. 

The true prior density, a weighed sum of Gaussian distributions, is denoted as
$$
    \rho(x) = \sum_{i = 1}^{N} w_{i}\frac{\exp \left(-\frac{1}{2}(x-\mu_{i})^{\intercal} {\Sigma_{i}}^{-1}(x-\mu_{i})\right)}{\sqrt{(2 \pi)^{2}|\Sigma_{i}|}}.
$$

Example 1 is a mixture of four Gaussians. The weights are $w_{1} = w_{2} = w_{3} = w_{4} = 0.25$. The means of the Gaussians are 
$$
\mu_{1} = [1, 0]^{\intercal}, \mu_{2} = [0, 1]^{\intercal}, \mu_{3} = [2, 2]^{\intercal}, \mu_{4} = [-2, -2]^{\intercal}
$$
and the covariance matrices are
$$
\Sigma_{1} = \Sigma_{2} = \Sigma_{3} = \Sigma_{4} = \begin{bmatrix}
1 & 0\\ 
0 & 1
\end{bmatrix}.
$$
$\theta(x)$ is chosen as a multivariate Guassian distribution $\mathcal{N}\left(\begin{bmatrix}
0\\ 
0
\end{bmatrix}, \begin{bmatrix}
4 & 0\\ 
0 & 4
\end{bmatrix}\right).$

\begin{figure}[ht]
  \subfloat{\includegraphics[width=1\textwidth]{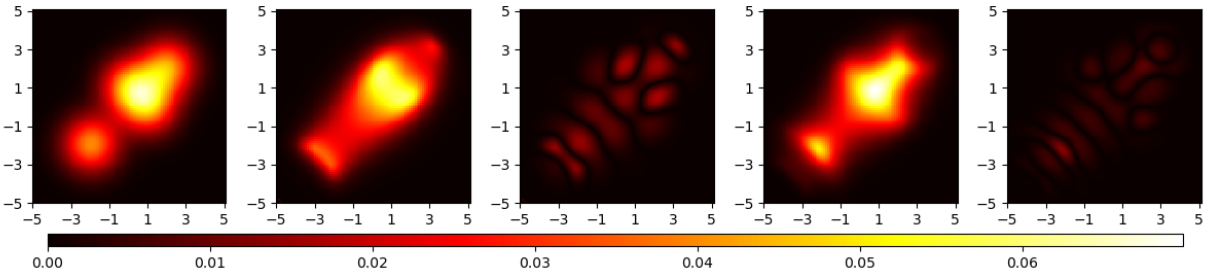}}
  \newline	
  \subfloat{\includegraphics[width=1\textwidth]{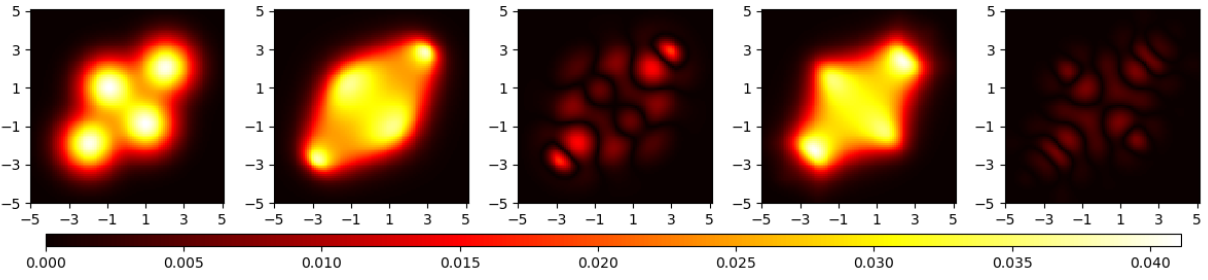}}
  \newline
  \subfloat{\includegraphics[width=1\textwidth]{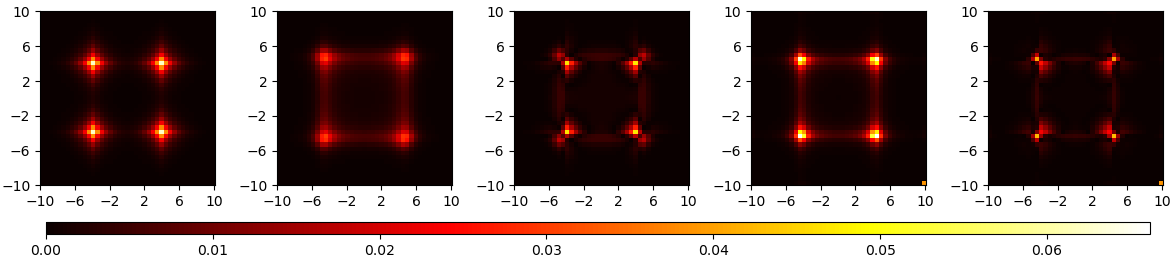}}
  \newline
  \subfloat{\includegraphics[width=1\textwidth]{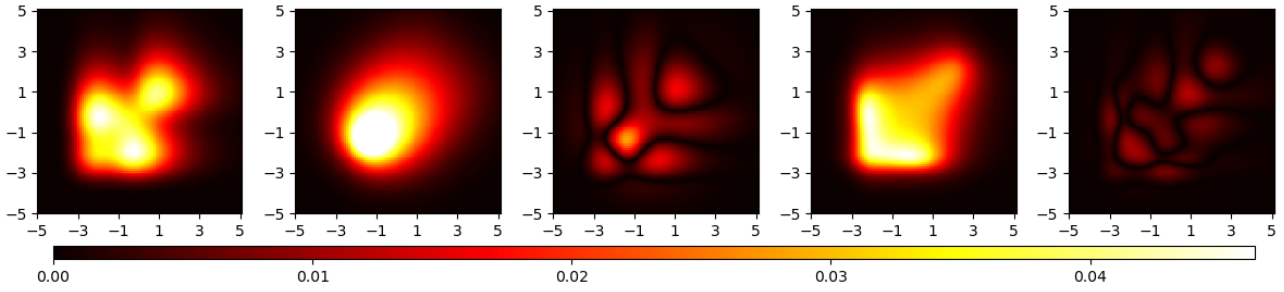}}
  \newline 	
  \subfloat{\includegraphics[width=1\textwidth]{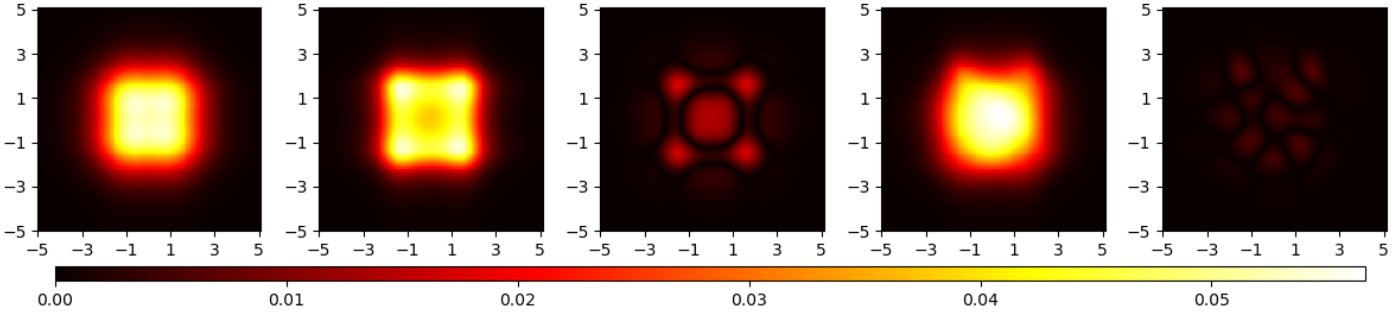}}
\caption{Simulation results of prior density estimation. Each row is the results of an example. The first column is the true prior density. The second and fourth columns are the density estimates with order $4$ and $6$ density surrogates respectively. The third and fifth columns are the absolute values of error of the density estimates.}
\label{figure4}
\end{figure}

The simulation results of Example 1 is shown in the first row of Figure \ref{figure4}. For each example the true prior is shown in the first column. We first choose the highest order of each variable $x_{i}$ in the polynomial $q(x, \Lambda)$ to be $4$. The density estimate $\hat{\rho}(x)$ is shown in the second column and the third column is the absolute value of error of the density estimates, i.e., $|\rho - \hat{\rho}|$, and $\max |\rho - \hat{\rho}| = 0.0175$. We then choose the highest order of each variable $x_{i}$ in the polynomial $q(x, \Lambda)$ to be $6$. The density estimates and the corresponding absolute value of error is given in the fourth and fifth figures of the first row. $\max |\rho - \hat{\rho}| = 0.0143$.

By the simulation results of the first example, we note that our proposed parametrization is able to approximate the prior density well with power moments up to order $4$ and order $6$. With the increase of order, a better estimation result is obtained. This simulation also reveals the fact that the power moment terms are a more compact representation of the density as compared to the discrete data points. Only $(2 \cdot 2+1)^{2} = 25$ for order $4$, and $(2 \cdot 3+1)^{2} = 49$ power moment terms are used for estimating the prior density, i.e., only $25$ and $49$ parameters are used to characterize the density for order $4$ and order $6$ respectively. However it is not quite possible for $25$ or $49$ discrete particles to represent a density with such a high accuracy.

As to verify the performance of the proposed density surrogate for estimating different mixture of Gaussian densities, we simulate on the following example.

Example 2 is chosen as a mixture of four Gaussians. The weights are $w_{1} = w_{2} = w_{3} = w_{4} = 0.25$. The means of the Gaussians are 
$$
\mu_{1} = [1, -1]^{\intercal}, \mu_{2} = [-1, 1]^{\intercal}, \mu_{3} = [2, 2]^{\intercal}, \mu_{4} = [-2, -2]^{\intercal}
$$
and the covariance matrices are
$$
\Sigma_{1} = \Sigma_{2} = \Sigma_{3} = \Sigma_{4} = \begin{bmatrix}
1 & 0\\ 
0 & 1
\end{bmatrix}.
$$

$\theta(x)$ is a Gaussian distribution, which is the same as that in Example 1. The highest orders of the polynomial $q(x)$ are $4$ and $6$. The simulation results are given in the second row of Figure \ref{figure4}. We note that four modes(peaks) are well estimated by the density surrogate. The absolute value of error $\max |\rho - \hat{\rho}|$ are $0.0191$ and $0.0071$ respectively for order $4$ and $6$.

In the previous two examples, the performance of our proposed density surrogate is validated by simulations on mixtures of Gaussian distributions. In the following three examples, we simulate more complicated mixtures of densities, which have not been considered in the previous results.

Example 3 is chosen as a mixture of four multivariate generalized asymmetric Laplace distributions. A multivariate generalized asymmetric Laplace (GAL) distribution is defined as follows \cite{kozubowski2013multivariate}. If the matrix $\Sigma_{i}$ is positive-definite, the GAL distribution is truly $d$-dimensional and has a probability density function of the form
$$
\begin{aligned}
    \rho_{i}(x)& =\frac{2 \exp \left(\mu_{i}^{\prime} \Sigma_{i}^{-1} x\right)}{(2 \pi)^{d / 2} \Gamma(s)|\Sigma_{i}|^{1 / 2}}\left(\frac{Q_{i}(x)}{C_{i}(\Sigma_{i}, \mu_{i})}\right)^{s-d / 2}\\
    & \cdot K_{s-d / 2}(Q_{i}(x) C_{i}(\Sigma_{i}, \mu_{i})),
\end{aligned}
$$
where $K_{\lambda}(\cdot)$ is the modified Bessel function with index $\lambda$ \cite{olver2010digital} and
$$
Q_{i}(x)=\sqrt{x^{\prime} \Sigma_{i}^{-1} x}, \quad C_{i}(\Sigma_{i}, \mu_{i})=\sqrt{2+\mu_{i}^{\prime} \Sigma_{i}^{-1} \mu_{i}}.
$$

We simulate $\rho(x) = \sum_{i = 1}^{4}w_{i}\rho_{i}(x)$ with $w_{1} = w_{2} = w_{3} = w_{4} = 0.25$. The means of each GAL distribution $\rho_{i}(x)$ are respectively
$$
\mu_{1} = [4, 4]^{\intercal}, \mu_{2} = [4, -4]^{\intercal}, \mu_{3} = [-4, 4]^{\intercal}, \mu_{4} = [-4, -4]^{\intercal}
$$
and the covariance matrices are
$$
\Sigma_{1} = \Sigma_{2} = \Sigma_{3} = \Sigma_{4} = \begin{bmatrix}
1 & 0\\ 
0 & 1
\end{bmatrix}.
$$

$\theta(x)$ is chosen as a multivariate Guassian distribution $\mathcal{N}\left(\begin{bmatrix}
0\\ 
0
\end{bmatrix}, \begin{bmatrix}
9 & 0\\ 
0 & 9
\end{bmatrix}\right).$ The highest orders of the polynomial $q(x)$ are $4$ and $6$. The simulation results are given in the third row of Figure \ref{figure4}. We note that both two estimates are able to approximate the four sharp peaks with no prior knowledge of the prior density, i.e., the prior density $\theta(x)$ is chosen as a multivariate Gaussian distribution. By the simulation results, we note that with higher order power moments used, the estimated peaks tend to be closer to the true ones. 

Example 4 is chosen as a mixture of four multivariate Gumbel distributions, where the modes are not clearly separated. The probability density function we consider for simulation is
$$
     \rho(x) = w_{i}\sum_{i = 1}^{4}\prod_{j = 1}^{2} \exp\left(-\left(x_{j} - \mu_{i, j} + \exp\left(- (x_{j} - \mu_{i, j}) \right)\right)\right)
$$
where $\mu_{i, j}$ denotes the $j_{\text{th}}$ element of the mean vector $\mu_{i}$.
The weights are $w_{1} = w_{2} = w_{3} = w_{4} = 0.25$, and the mean vectors are
$$
\mu_{1} = [1, 1]^{\intercal}, \mu_{2} = [-2, 0]^{\intercal}, \mu_{3} = [0, -2]^{\intercal}, \mu_{4} = [-2, -2]^{\intercal}.
$$

$\theta(x)$ is a Gaussian distribution, which is the same as that in Example 1. The highest orders of the polynomial $q(x)$ are $4$ and $6$. The simulation results are given in the third row of Figure \ref{figure4}. With power moments up to order $4$, the estimates is not able to approximate the four modes of the prior density well. With more power moments, the density estimation is satisfactory. The absolute value of error $\max |\rho - \hat{\rho}|$ are $0.0283$ and $0.0118$ respectively for order $4$ and $6$. 

Heavy-tailed filtering problem is drawing more attentions in the control community in the recent years due to its applications in intelligent vehicles and underwater robots \cite{hanzon2001state, zhu2021novel, bai2020novel}, which are working in environments with possibly more outlier observations. The proposed non-Gaussian Bayesian filter makes it feasible for us to treat this problem by choosing $\theta(x)$ as a heavy-tailed distribution. In the following example, we simulate mixtures of student-t distributions. 

Example 5 is chosen as a mixture of four student-t distributions. The probability density function we consider for simulation is
$$
    \rho(x)= \sum_{i = 1}^{4}\prod_{j = 1}^{2} \frac{\Gamma\left(\frac{\nu+1}{2}\right)}{\sqrt{\nu \pi} \Gamma\left(\frac{\nu}{2}\right)}\left(1+\frac{(x_{j} - \mu_{i, j})^{2}}{\nu}\right)^{-(\nu+1) / 2}
$$
The mean vectors are chosen as
$$
\mu_{1} = [1, 1]^{\intercal}, \mu_{2} = [1, -1]^{\intercal}, \mu_{3} = [-1, 1]^{\intercal}, \mu_{4} = [-1, -1]^{\intercal}.
$$

We note that the power moments up to order $\nu-1$ exist and are finite, for $\nu \in \mathbb{N}_{0}$. So then we choose $\nu = 8$ to ensure that the density surrogates with moments up to order $4$ and $6$ both exist. $\theta(x)$ is chosen as $\mathcal{C}(0, 3)$, where $\mathcal{C}$ denotes the Cauchy distribution. The highest order of the polynomial $q(x)$ is $4$ and $6$. The simulation results are given in the fifth row of Figure \ref{figure4}. With power moments up to order $4$, the estimates is not able to approximate the four modes which overlap with each other. And the tail is estimated to be narrower.  However with more power moments, the overlapped modes are well approximated. And the tail is well characterized by the density surrogate. The absolute value of error $\max |\rho - \hat{\rho}|$ are $0.0283$ and $0.0118$ respectively for order $4$ and $6$.

In conclusion, the $5$ examples show the performance of our proposed Bayesian filter in estimating the intractable prior density which can be either light-tailed (sub-Gaussian) or heavy-tailed. The proposed density surrogate doesn't require prior knowledge of either the number of modes or the feasible function classes of the intractable prior density, which is a clear and significant advantage for real applications.

\onecolumn

\twocolumn
\subsection{A robot localization problem}
\begin{figure}[h]
\centering
\includegraphics[scale=0.28]{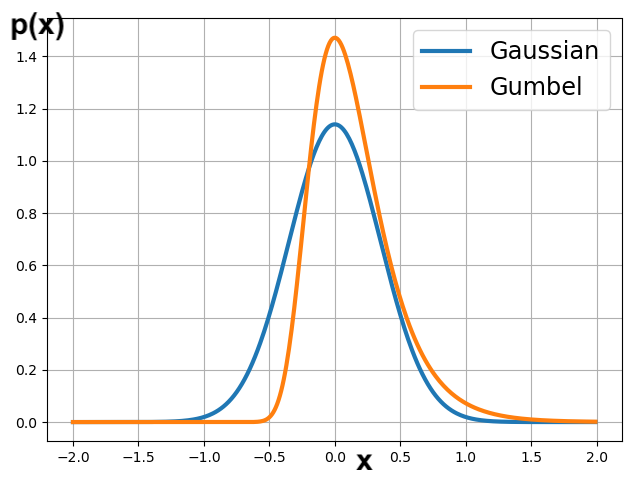}
\centering
\caption{Probability density functions of the Gaussian and the Gumbel distributions.}
\label{figC1}
\end{figure}

In this subsection, we focus on a robot localization challenge involving a sensor tasked with measuring the distances between the robot and predefined landmarks. We approximate a  small area of the ground as a Euclidean plane equipped with two perpendicular coordinate axes. The robot's position with respect to these coordinates at time step $k$ is denoted as $\{x(k), y(k)\}$. Furthermore, we denote the positions of $L$ landmarks as $\{\check{x}_{1}, \check{y}_{1}\}, \cdots, \{\check{x}_{L}, \check{y}_{L}$\}. We note that there need to be at least two landmarks, or the location of the robot is not completely observable. In this localization task, the robot is incrementally moved one unit along the positive $x$ and one unit along the positive $y$ directions. The robot's controls are not perfect so it will not move exactly as commanded. Hence we need to consider noise in the particle's movements to have a reasonable chance of capturing the actual movement of the robot. In this simulation, we assume the bearing angle, of which the true value is $\pi/4$, to have an additive Gaussian noise $\mathcal{N}(0, 0.2^{2})$. In addition, we assume the moving distance, of which the true value is $\sqrt{2}$, to be corrupted with an additive Gaussian noise $\mathcal{N}(0, 0.05^{2})$. Besides, the distance observation of each landmark is corrupted with an additive noise. The robot localization problem is then formulated as follows. Denote the distance of the robot to each known landmark $i$ as $z_{i}(k)$. The system and observation equations read
$$
\begin{bmatrix}
x(k+1)\\ 
y(k+1)
\end{bmatrix} = \begin{bmatrix}
x(k)+1\\ 
y(k)+1 
\end{bmatrix} + \begin{bmatrix}
w_{1}(k)\\ 
w_{2}(k)
\end{bmatrix}
$$
and
$$
\begin{bmatrix}
z_{1}(k)\\ 
\vdots\\ 
z_{L}(k)
\end{bmatrix} = \begin{bmatrix}
\sqrt{\left( x(k)-\check{x}_{1} \right )^{2} + \left( y(k)-\check{y}_{1} \right )^{2}}\\ 
\vdots\\ 
\sqrt{\left( x(k)-\check{x}_{L} \right )^{2} + \left( y(k)-\check{y}_{L} \right )^{2}}
\end{bmatrix} + \begin{bmatrix}
v_{1}(k)\\ 
\vdots\\ 
v_{L}(k)
\end{bmatrix}
$$
respectively. We assume $w_{1}(k), w_{2}(k)$ to follow the Gaussian distributions $\mathcal{N}(0, 0.1^{2})$ which takes the error of controlling the robot into consideration. In the previous results, the noise $v_{1}(k), \cdots, v_{L}(k)$ are always assumed to be Gaussian. However, the Gaussian distribution is not an ideal model to characterize the observation error. The reason is that when the landmark is closer, the robot is more likely to detect it and the distance observation is more accurate. However when the landmark is farther, it is more probable that the distance observation is biased more severely from the true one. In this example, we propose to use the right-skewed Gumbel distribution as an example to characterize this property, and to validate the proposed filtering scheme. A comparison of the probability density functions of the Gumbel and Gaussian distributions are given in Figure \ref{figC1}, where the probability density function of the Gumbel distribution is
\begin{equation}
p_{v_{i}}(x) = 4 e^{-4x-e^{-4x}}.
\label{pv1}
\end{equation}
The Gaussian distribution has identical mean and variance as the Gumbel distribution, of which the probability density function reads
\begin{equation}
p_{v_{i}}(x) = \frac{1}{\sqrt{2\pi} \cdot 0.35}e^{\frac{x^{2}}{2 \cdot 0.35^{2}}}.
\label{pv2}
\end{equation}

When $v_{i} > 0$, we note that it is more possible for $v_{i}$, which follows the Gumbel distribution, to be biased from $0$. However, when $v_{i} < 0$, the probability value converges to zero more quickly with the decrease of $v_{i}$, compared with the Gaussian distribution. Therefore, the Gumbel distribution characterizes the property well.

However, the asymmetric Gumbel distribution, which acts as the model of observation noise in this localization task, causes severe problem in selecting a proper stochastic filter. To our knowledge, except for the multivariate filter based on power moments proposed (MF) in this paper, only the particle filter (PF) is feasible of performing this task. We adopt a typical sampling-importance resampling (SIR) filter \cite{chen2003bayesian} in this simulation. Since the observation equation is a nonlinear equation of the system state, we also simulate the problem using the unscented Kalman filter (UKF). However, the UKF is not able to treat the Gumbel observation noise. In this case, we adopt the Gaussian distribution in \eqref{pv2} as a substitute of \eqref{pv1} as the observation noise for filtering. 

\begin{figure}[h]
\centering
\includegraphics[scale=0.32]{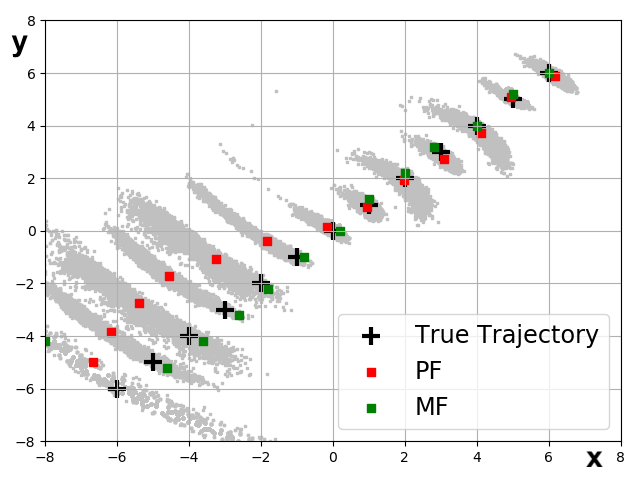}
\centering
\caption{A sample localization process. The black crosses represent the true trajectory of the robot. The red and green dots represent the location estimates by the particle filter and our proposed filter by multivariate power moments respectively. The gray dots represent the particles of the particle filter at each time step.}
\label{figC2}
\end{figure}

\begin{figure}[h]
\centering
\includegraphics[scale=0.32]{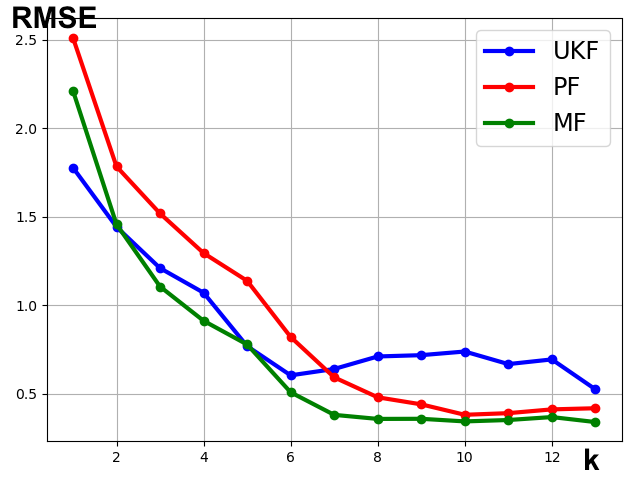}
\centering
\caption{RMSE as a function of time step $k$ of $50$ Monte-Carlo simulations for the unscented Kalman filter (UKF), the particle filter (PF) and the proposed filter by multivariate power moments (MF).}
\label{figC3}
\end{figure}

The initialization of the three filters are as follows. The robot starts moving from $[-6, -6]^{\intercal}$. The four landmarks are located at $[-1, 2]^{\intercal}, [5, 10]^{\intercal}, [12, 14]^{\intercal}, [18, 21]^{\intercal}$. The initial states $[x(0), y(0)]^{\intercal}$ of the MF and UKF are drawn from the Gaussian distribution $\mathcal{N}\left(\begin{bmatrix}
-6\\ 
-6
\end{bmatrix}, \begin{bmatrix}
2^{2} & 0 \\ 
0 & 2^{2} 
\end{bmatrix}\right)$. The states of the $5000$ initial particles of the PF are drawn from the Gaussian distribution $\mathcal{N}\left(\begin{bmatrix}
-6\\ 
-6
\end{bmatrix}, \begin{bmatrix}
5^{2} & 0 \\ 
0 & 5^{2} 
\end{bmatrix}\right)$, where the variances are relatively larger to cover a wider range of possible locations. The additive noise of the distance observation follows the distribution in \eqref{pv1}. We use multivariate power moments up to order $4$ to estimate the density surrogates in the implementation of MF. A sample localization process is given in Figure \ref{figC2}, where the estimation results of PF and MF are given. We note that the location estimates of the MF converge to the true locations. The states of the particles of the PF also converge to the true locations. In Figure \ref{figC3}, the root mean square error (RMSE) curves of the MF, PF and UKF of $50$ Monte-Carlo simulations are given. We note that when the state estimate converges, the RMSE errors of MF and PF are close, with the one of MF slightly smaller. However, the RMSE of the UKF is significantly larger than those of the MF and PF. It is inevitable since a Gaussian distribution is used as an approximation of the true Gumbel distribution. Moreover, the convergence of the RMSE for the UKF cannot be assured based on the curve.

From the perspective of RMSE, the MF isn't superior than the PF. However, a significant disadvantage of the Particle filter is that it needs to store massive data. For example in this simulation, we need $3$ parameters for each particle, including its two-dimensional position and the weight. Hence we need $15000$ parameters to characterize the density of the system state. In this simulation, the system state is only $2$ dimensional. With the increase of the dimension, the parameters required increases exponentially, which we may not be able to handle. While for the MF, we only need $(4+1)^{2} = 25$ parameters for this task, which is a much more compact way for the representation of the density function. 

Moreover, since a convex optimization is requred in each filtering step, we need to take the time consumption into consideration. For each iteration of filtering in this example, the average processing time is 6.8 seconds on a 2.5 GHz Intel Core i7 CPU. It is a relatively long time compared to the processing time of the PF. However, it is not quite a long period of time for applications where the processing time is not very sensitive. Moreover, each optimization problem is a convex one with the existence and uniqueness of solution proved. The processing time can be decreased by designing a strategy determining the step length of the gradient methods.

\section{Conclusion}

A multivariate non-Gaussian
Bayesian filter with the state estimation parameterized as an analytic function is proposed in this paper, where the distribution of the observation noise can be a Lebesgue integrable function, and that of the system noise can be either a probability mass function or a Lebesgue integrable function. It is significant that the proposed algorithm is able to estimate the density functions without prior knowledge of the density of the state $x_{t}$, e.g. the number of modes and the feasible function class. It is not required to store massive estimates of the state at discrete points. We prove that the estimated power moments are asymptotically unbiased, approximately the true ones throughout the filtering process given a sufficiently large $n$. The existence of solution to the multidimensional Hamburger moment problem is established, and a novel Positivstellensatz is proposed to ensure the positiveness of the density surrogate, which also serves as a new result to the moment problem. The parameters of the proposed parametrization can be obtained by a convex optimization scheme. The solution to this problem is proved to exist and be unique by proving that the map from the parameters to the power moments is homeomorphic. Upper bounds of the state estimate are also proposed. In the simulations, we first estimate mixtures of different types of multivariate density functions using power moments, including Gaussian, Laplacian, Gumbel and student-t. The simulation results on the mixture of student-t distributions validates the ability of the proposed algorithm to treat the heavy-tailed filtering problem, which is a current key problem of stochastic filtering. We also simulate the algorithm on a robot localization problem, with a comparison to the particle filter and the unscented Kalman filter. The simulation results reveal the potential of the proposed filter in real engineering applications.

\bibliographystyle{plain}
\bibliography{main}

\begin{IEEEbiography}[{\includegraphics[width=1in,height=1.25in,clip,keepaspectratio]{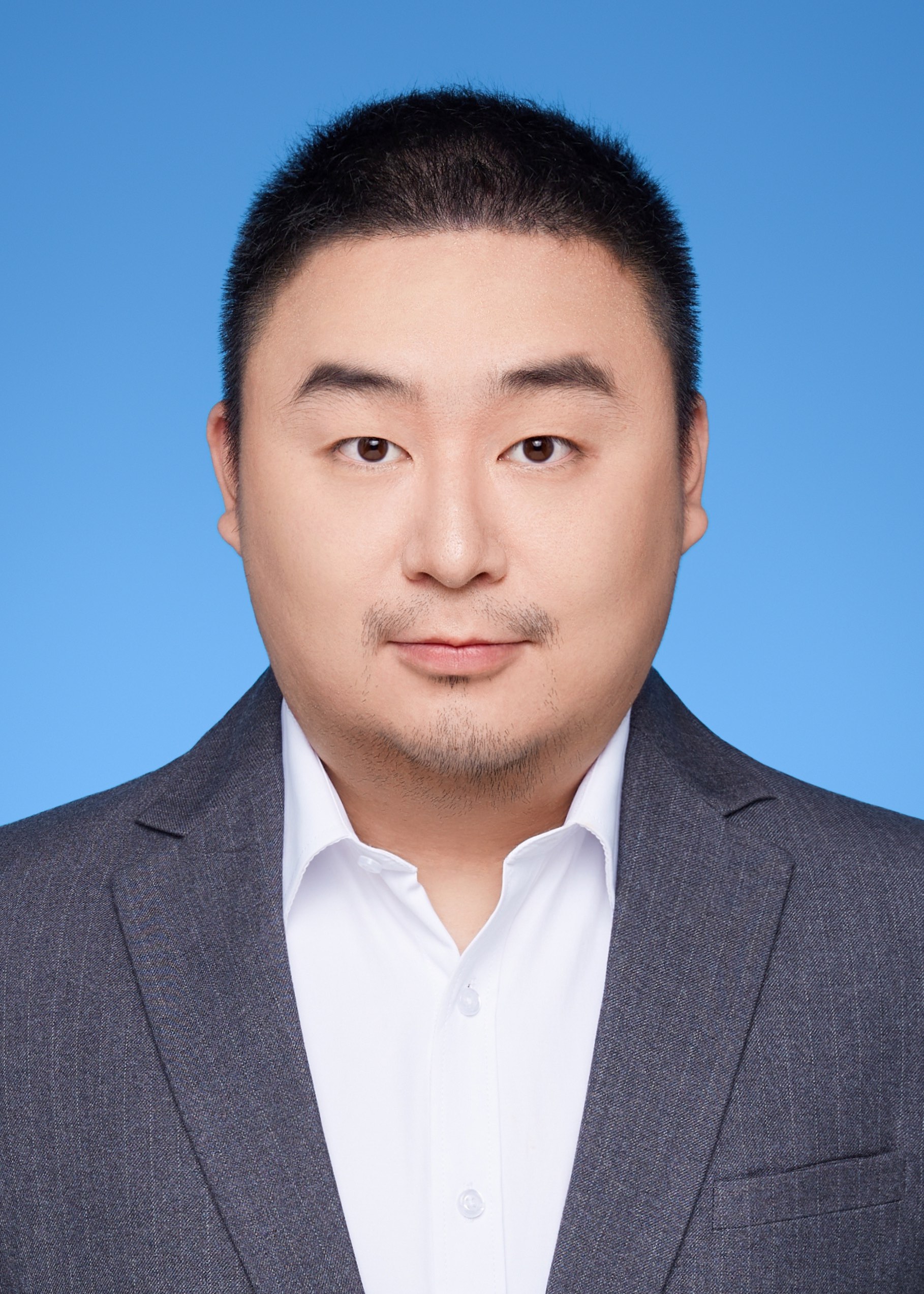}}]{Guangyu Wu} (S'22) received the B.E. degree from Northwestern Polytechnical University, Xi’an, China, in 2013, and two M.S. degrees, one in control science and engineering from Shanghai Jiao Tong University, Shanghai, China, in 2016, and the other in electrical engineering from the University of Notre Dame, South Bend, USA, in 2018. He is pursuing the Ph.D. degree at Shanghai Jiao Tong University.

\end{IEEEbiography}

\begin{IEEEbiography}[{\includegraphics[width=1in,height=1.25in,clip,keepaspectratio]{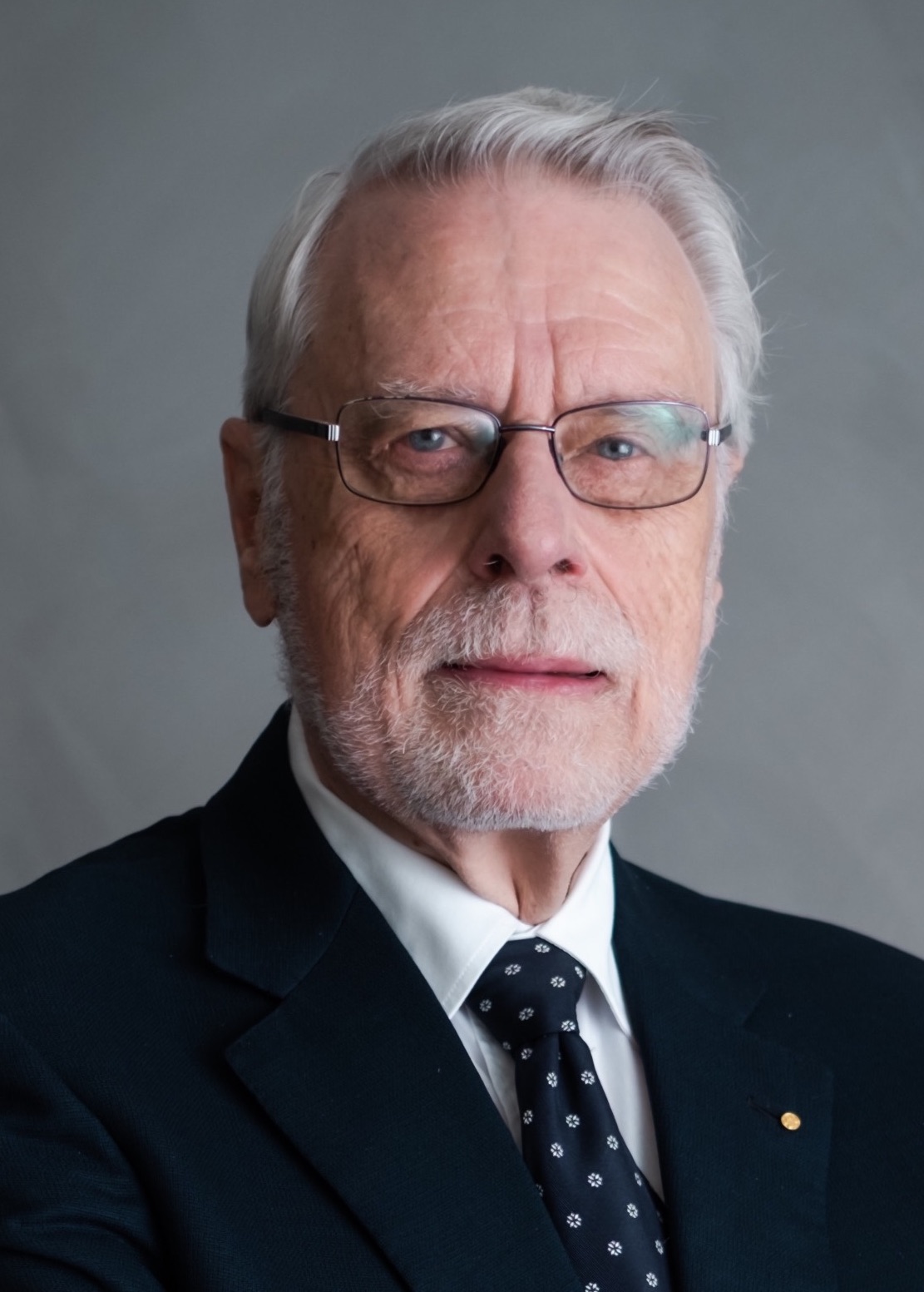}}]{Anders Lindquist} (M’77–SM’86–F’89–LF’10) received the Ph.D. degree in optimization and systems theory from the Royal Institute of Technology (KTH), Stockholm, Sweden, in 1972, and an honorary doctorate (Doctor Scientiarum Honoris Causa) from Technion (Israel Institute of Technology) in 2010, and Doctor Jubilaris from KTH in 2022. 

He is currently a Zhiyuan Chair Professor at Shanghai Jiao Tong University, China, and Professor Emeritus at KTH, Stockholm, Sweden. Before that he had a full academic career in the United States, after which he was appointed to the Chair of Optimization and Systems at KTH.
Dr. Lindquist is a Member of the Royal Swedish Academy of Engineering Sciences, a Foreign Member of the Chinese Academy of Sciences, a Foreign Member of the Russian Academy of Natural Sciences, a Member of Academia Europaea (Academy of Europe), an Honorary Member the Hungarian Operations Research Society, a Life Fellow of IEEE, a Fellow of SIAM, and a Fellow of IFAC. He received the 2003 George S. Axelby Outstanding Paper Award, the 2009 Reid Prize in Mathematics from SIAM, and the 2020 IEEE Control Systems Award, the
IEEE field award in Systems and Control.
\end{IEEEbiography}

\end{document}